\documentclass{aa}  

\usepackage{lscape}
\usepackage{graphicx}
\usepackage{float}
\usepackage{txfonts}
\begin{document}

   \title{Probing the ejecta of evolved massive stars in transition}

   \subtitle{A VLT/SINFONI $K$-band survey\thanks{Based on observations at the European Southern Observatory, 
Paranal, under program IDs 384.D-1078(A) and 088.D-0442(B).}}

   \author{M. E. Oksala \inst{1}
          \and
          M. Kraus \inst{1}
          \and
          L. S. Cidale \inst{2}\fnmsep\inst{3}
          \and
          M. F. Muratore \inst{2}\fnmsep\inst{3}
          \and
          M. Borges Fernandes \inst{4}}

   \institute{Astronomick\'{y} \'{u}stav, Akademie v\v{e}d \v{C}esk\'{e} republiky, Fri\v{c}ova 298, 251 65 Ond\v{r}ejov, Czech Republic\\
              \email{oksala@sunstel.asu.cas.cz}
             \and
             Departamento de Espectroscop\'ia Estelar, Facultad de Ciencias Astron\'{o}micas y Geof\'{i}sicas, Universidad Nacional de La Plata (UNLP), Paseo del Bosque S/N, 1900 La Plata, Argentina
             \and 
             Instituto de Astrof\'{i}sica La Plata, CCT La Plata, CONICET, Paseo del Bosque S/N, 1900 La Plata, Argentina
             \and
             Observat\'{o}rio Nacional, Rua General Jos\'{e} Cristino, 77 S\~{a}o Cristov\~{a}o, 20921-400, Rio de Janeiro, Brazil
             }

   \date{Received ; accepted}

% \abstract{}{}{}{}{} 
% 5 {} token are mandatory
 
  \abstract
{Massive evolved stars in transition phases, such as Luminous Blue Variables (LBVs),
B[e] Supergiants (B[e]SGs), and Yellow Hypergiants (YHGs), are not well understood,
and yet crucial steps in determining accurate stellar and galactic evolution models.
The circumstellar environments of these stars reveal their mass-loss history, identifying 
clues to both their individual evolutionary status and the connection between objects of different phases.
Here we present a survey of 25 such evolved massive stars (16 B[e]SGs, 6 LBVs, 2 YHGs, and 1 Peculiar Oe star), 
observed in the $K$-band with the Spectrograph for INtegral Field Observation in the Near-Infrared 
(SINFONI; R = 4500) on the ESO VLT UT4 8~m telescope.  The sample can be split into two 
categories based on spectral morphology: one group includes all of the B[e]SGs, the 
Peculiar Oe star, and two of the LBVs, while the other includes the YHGs and the rest
of the LBVs.  The difference in LBV spectral appearance is due to some objects being in a quiescent 
phase and some objects being in an active or outburst phase. CO emission features are found
in 13 of our targets, with first time detections for MWC 137, LHA 120-S 35, and LHA 115-S 65.
From model fits to the CO band heads, the emitting regions appear to be detached from the stellar 
surface.  Each star with $^{12}$CO features also shows $^{13}$CO emission, signaling an evolved nature.
Based on the level of $^{13}$C enrichment, we conclude that many of the B[e]SGs are likely in a pre-Red 
Supergiant phase of their evolution. There appears to be a lower luminosity limit of 
$\log$ L/L$_{\odot}$ = 5.0 below which CO is not detected.  The lack of CO features in several high 
luminosity B[e]SGs and variability in others suggests that they may in fact be LBV candidates, 
strengthening the connection between these two very similar transition phases.
}


   \maketitle

\section{Introduction}

%\begin{landscape}
\begin{sidewaystable*}
\caption{Literature stellar parameters.  }             
\label{tbl-1}      
\centering          
\begin{tabular}{lccccccccccl} 
\hline\hline      
Object ID   & Alternate IDs & RA(J2000) & 
Dec(J2000) & Spectral & Object & $K_{s}$  & $\log$ T$_{\rm{eff}}$ &
$\log$ L & E(B-V) & $\varv_{\rm{sys}}$ & Ref. \\
 &  &  &  & Type  & Type & (mag) &  (K) & (L$_{\odot}$)  &  & (km s$^{-1}$) &   \\
\hline
\noalign{\smallskip}
\multicolumn{11}{c}{Small Magellanic Cloud} \\
\noalign{\smallskip}
\hline
LHA 115-S 6  & R4  &  00:46:54.4 & -73:08:37.3 & B0   &  B[e]SG  & 11.025 &  4.43      & 5.00   & 0.07  &  152  & 1,2  \\
LHA 115-S 18 &  &  00:54:08.9 & -72:41:46.1 & B0   &  B[e]SG  & 11.109 &  4.40      & 5.66   & 0.26  &  146  & 3    \\
LHA 115-S 52  &  HD 6884, R40  &  01:07:17.6 & -72:28:06.2 & --  &  LBV     & 9.466  &  3.94      & 5.50   & 0.14  &  170  & 4 \\
LHA 115-S 65  & R50  &  01:44:02.9 & -74:40:53.5 & B2-3 &  B[e]SG  & 9.724  &  4.23      & 5.69   & 0.15  &  191  & 1  \\
\hline
\noalign{\smallskip}
\multicolumn{11}{c}{Large Magellanic Cloud} \\
\noalign{\smallskip}
\hline
HD 269723    & R117  &  05:32:24.3 & -67:41:57   & G4 Ia&  YHG     & 7.689  &  3.67      & 5.50   & 0.19  &  332  & 4  \\
HD 269953     & R150  &  05:40:11.5 & -69:40:07.9 & G0 I &  YHG     & 8.021  &  3.69      & 5.59   & 0.11  &  242  & 4 \\
LHA 120-S 12  & &  04:57:36.3 & -67:47:39.9 & B0.5 &  B[e]SG  & 10.229 &  4.36      & 5.33   & 0.25  &  290  & 1   \\
LHA 120-S 22  & HD 34664 &  05:13:52.4 & -67:26:56.4 & B0.5 &  B[e]SG  & 8.462  &  4.36      & 5.78   & 0.25  &  330  & 1  \\
LHA 120-S 35  &  &  05:27:17.7 & -66:22:05.1 &B1 Iab&  B[e]SG  & 10.801 &  4.34      & 5.20   & 0.06  &  308  & 5  \\
LHA 120-S 59  &  &  05:45:28.9 & -68:11:49.5 & B5 II&  B[e]SG  & 11.475 &  4.15      & 4.00   & 0.01  &  295  & 5  \\
LHA 120-S 73   & HD 268835, R66 &  04:56:46.7 & -69:50:27.5 &  B8I &  B[e]SG  & 8.855  &  4.08      & 5.45   & 0.12  &  261  & 1  \\
LHA 120-S 89  & HD 269217, R82 &  05:13:38.6 & -69:21:11.5 & B2-3 &  B[e]SG  & 9.803  &  4.27      & 5.41   & 0.20  &  236  & 1  \\
LHA 120-S 93  &  &  05:16:31.5 & -68:22:11.7 & B9Ib &  B[e]SG  & 10.519 &  4.00      & 4.70   & 0.20  &  288  & 5  \\
LHA 120-S 96  & S Dor, HD 35343, R88 &  05:18:13.7 & -69:15:03.3 & --  &  LBV     & 8.338  &  4.35/3.95 & 5.98   & 0.05  &  295  & 4  \\
LHA 120-S 116  & HD 269662, R110 &  05:30:51.2 & -69:03:01.7 & --  &  LBV     & 9.504  &  4.01/3.88 & 5.46   & 0.10  &  282  & 4  \\
LHA 120-S 124 & HD 37836, R123 &  05:35:15.9 & -69:40:40.7 & O8   &  Pec Oe  & 9.381  &  4.45      & 6.10   & 0.15\tablefootmark{a}  &  267  &  6 \\
LHA 120-S 127  & HD 37974, R126 &  05:36:25.4 & -69:22:58.2 & B0.5 &  B[e]SG  & 8.760  &  4.35      & 6.10   & 0.25  &  258  & 1  \\
LHA 120-S 128  & HD 269858, R127 &  05:36:43.3 & -69:29:50.5 & --  &  LBV     & 9.225  &  4.48/3.93 & 6.10   & 0.20  &  284  & 4  \\
LHA 120-S 134  & HD 38489 &  05:40:12.9 & -69:22:50.1 &  B0  &  B[e]SG  & 8.602  &  4.41      & 5.90   & 0.25  &  267  & 1 \\
LHA 120-S 137  & &  05:41:43.6 & -69:37:41.6 & B6 Ib&  B[e]SG  & 11.141 &  4.11      & 4.18   & 0.14  &  272  & 5  \\
LHA 120-S 155  & HD 269006, R71 &  05:02:06.5 & -71:20:16.6 & --  &  LBV     & 10.547 &  4.13/3.95 & 5.42   & 0.15  &  198  & 4  \\
\hline
\noalign{\smallskip}
\multicolumn{11}{c}{Milky Way Galaxy} \\
\noalign{\smallskip}
\hline
MWC 137      &  V1308 Ori &  06:18:45.5 & 15:16:51.46 & B0   &  B[e]SG  & 6.623  &  4.48      & 5.37   & 1.216 &   18  & 7 \\
GG Car        & HD 94878 &  10:55:58.7 & -60:23:34.2 & B1I  &  B[e]SG  & 4.973  &  4.36      & 5.41   & 0.51  &  -25  & 8  \\
Hen 3-298     & &  09:36:44.1 & -53:28:01.8 &$<$ B3&  B[e]SG  & 5.697  &  4.11      & 5.10   & 1.70  &   82  & 9  \\
WRAY 15-751   & V432 Car &  11:08:39.9 & -60:42:50.6 & --  &  LBV     & 6.751  &  4.48/3.9  & 5.80/5.4   & 1.8   & 100\tablefootmark{a}  &  10,11 \\
\hline
\end{tabular}
\tablefoot{$K_{s}$-band magnitudes taken from the 2MASS point source catalog \citep{Cutri}.
\tablefoottext{a}{Values unknown, average values used in data reduction.}}
\tablebib{(1)~\citet{Zickgraf86}; (2)~\citet{Zickgraf96}; (3)~\citet{Zickgraf89}; (4)~\citet{deJager}; 
(5)~\citet{Gummer}; (6)~\citet{McGregor88b}; (7)~\citet{EF98}; (8)~\citet{Marchiano}; (9)~\citet{Miro05}; 
(10)~\citet{Sterken}; (11)~\citet{Hu}. }
\end{sidewaystable*}
%\end{landscape}

The post-main sequence evolution of massive (M$_{\rm{ini}} \ge 8$ M$_{\odot}$) stars 
remains one of the most difficult and confusing problems facing both observation 
and theory.  Although there are well behaved evolved stars, following theoretical prescribed paths 
through the supergiant phases, much interest has been paid to the more perplexing transition stages.
The path from the end of the main sequence through to the point of 
supernova can be substantially different for stars of varying mass, metallicity, 
rotation, etc. \citep[see e.g.,][]{MM04, Meynet11}, which can result in inaccurate 
stellar evolutionary model predictions, and as a result inadequate galactic evolution models.
During the transition phases, large amounts of mass are released from the stellar 
surface by strong stellar winds and/or eruption events.  The ejected material is detected 
in the circumstellar environment as shells, rings, or disks, and conditions often 
can facilitate the formation of molecules and dust.  The mechanism triggering such 
mass ejections (i.e. rapid rotation, pulsation, other instabilities) is not well known and 
may be different for various phases.  Further complicating matters is the lack of information 
given in the optical spectra of these stars. Often, the circumstellar material masks the 
photospheric contribution with strong permitted and forbidden emission, making stellar parameters 
difficult to obtain.

Many of the transition phases, specifically Luminous Blue Variables (LBVs), B[e] Supergiants (B[e]SGs),
and Yellow Hypergiants (YHGs), are poorly understood and still remain quite puzzling.  
LBVs expel large amounts of mass via eruptions likely arising as the star encroaches
the Eddington luminosity limit \citep[for reviews see][]{HD94, Nota}.  
These stars exhibit a wide variety of physical signatures 
of this mass-loss, with observational variability presenting as their primary characteristic.  
LBVs become redder as they brighten, due to the simultaneous expansion and cooling of the photosphere as 
the star transitions visually from a hot supergiant in its quiescence (visual minimum) phase
to a cool supergiant during its outburst (visual maximum) period.  This sequence is referred to as the S-Dor cycle 
of LBVs and repeats on the timescale of months to years. The expansion of the photosphere is also
referred to as the formation of a ``pseudo-photosphere'', as described by \citet{Leitherer85}.  
Occasionally, an LBV may experience a more intense, longer
mass-loss period, or large eruption (e.g., $\eta$ Carinae, HD~5980).  This type of event typically occurs on timescales of a thousand years.
The progenitors of LBVs are very massive, very luminous stars, however the physical state of these stars 
prior to or post LBV classification is unknown.  B[e]SGs show various observational 
signatures of a dense circumstellar disk or ring, including narrow forbidden emission lines, 
molecular emission, and evidence for dust \citep[for a general review see][]{Lamers98}.  
\citet{Zickgraf85} proposed a disk formation mechanism 
for B[e] stars, in which rapid rotation produces a hybrid wind consisting of a fast line-driven polar component
and a slow, high density equatorial component \citep[with further study by, e.g.,][]{Lamers91,Pelupessy,Kraus03,Cure04,
CR04,Cure05,Kraus06}.  While the spectral features certainly support this process,
rotational speeds have been determined for only a few B[e]SGs including LHA 115-S 23 \citep{Kraus08}, 
LHA 115-S 65 \citep{Zickgraf00,Kraus10}, LHA 120-S 93 \citep{Gummer}, and LHA 120-S 73 \citep{Zickgraf06}. 
Moreover, several recent studies show that this mechanism may not be a feasible explanation, based on the 
detection of detached rings/disks \citep{Kraus10,Kastner10,Liermann}.
YHGs are suggested to be cool, post-Red Supergiant (post-RSG) objects surrounded by dense shells of material expelled, 
presumably, by pulsational instabilities
\citep{deJager}.  These objects could be possible progenitors of LBVs and/or B[e]SGs, although their involvement
in the evolution of massive stars remains unclear.  To make the evolutionary picture even more ambiguous, each of 
these distinct stages occupy similar space in the Hertzsprung-Russell Diagram (HRD), as shown in Fig.~4 of \citet{Aret}.
Currently, stellar evolution codes do not predict any of these transition phases.

In the circumstellar regions of massive stars, strong radiation-driven stellar winds 
should prevent the formation of molecules and dust. 
However, in a disk structure, CO molecules can shield themselves
from the destructive stellar radiation field, and attain a high enough density for CO molecular emission to be observed. 
The seminal work of \citet{McGregor88a}, \citet{McGregor88b}, and \citet{McGregor89} confirmed 
this for certain evolved emission line stars.
The existence of CO band head emission is only possible if certain physical features of the material are met.  
At warm temperatures ($\sim$2000-5000 K) and high densities (>$10^{10}$ cm$^{-3}$), molecules are predicted to be 
abundant and sufficiently excited to generate spectral band head features.  
These types of conditions are also found in the circumstellar environments of  pre-main sequence 
objects such as Herbig Ae/Be stars or Herbig Ae/B[e] stars, which makes it sometimes difficult to assign a star the 
proper evolutionary stage.

To distinguish the true evolutionary stage of these stars, \citet{Kraus09} and \citet{Liermann} 
have demonstrated that CO emission features are not only useful to evaluate the physical conditions 
of the circumstellar material, but also to determine the presence of enrichment of the carbon isotope $^{13}$C, 
brought to the surface as a consequence of chemical reactions in the core and mixing processes.
In the ISM, the ratio of $^{12}$C to $^{13}$C is approximately 90, which is the value used as the initial 
ratio in stellar evolution calculations \citep[e.g.,][]{Ekstrom}.  By the end of the star's main sequence 
lifetime it should have decreased to 20 (as discussed in Sect.~\ref{statusBe}).  Because many of 
these stars are shrouded by their circumstellar environment, it would be difficult to determine this measurement. 
However, if the ejected material is a product of post-main sequence events \citep{Kraus09}, 
the circumstellar matter seen in the spectrum of these stars should accurately reflect the composition of the stellar
surface and can be used as a diagnosis tool.

In this paper, we present the results of our medium resolution (R=4500) $K$-band 
survey of evolved massive stars: B[e]SGs, LBVs, YHGs and a Peculiar Oe star,
to extract information about their circumstellar material, and to determine their 
possible evolutionary connections. Section 2 presents the sample, the observations,
and the data reduction process.  Additionally, this section presents the 
observed spectral morphology results.
Spectral models are explained and results reported in Section 3.  In Section 4, 
we discuss the specific outcomes of this study, 
and its possible implications for understanding massive star evolution.  
Section 5 summarizes our conclusions and states suggestions for future work.

\section{Observations and Spectral Morphology}

\begin{figure*}
\centering
\begin{tabular}{cc}
\includegraphics[width=80mm]{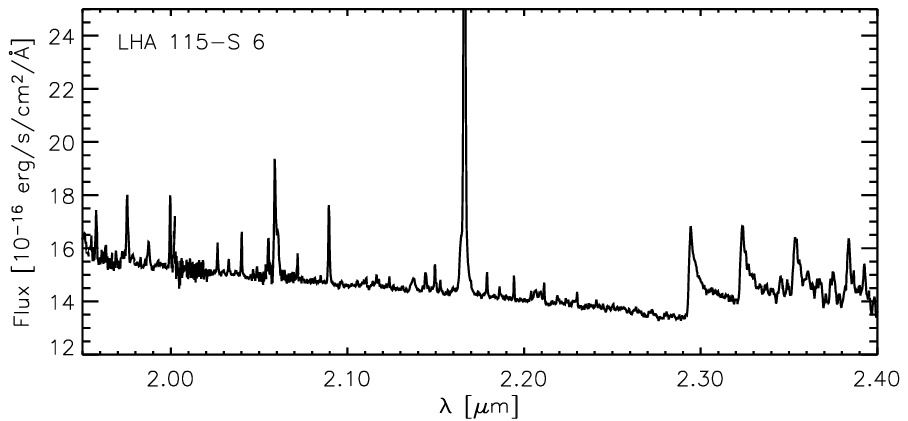} &
\includegraphics[width=80mm]{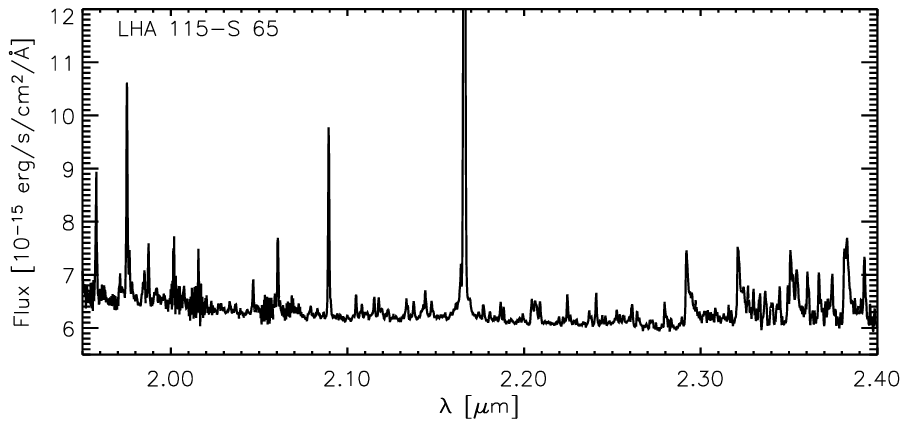} \\
\includegraphics[width=80mm]{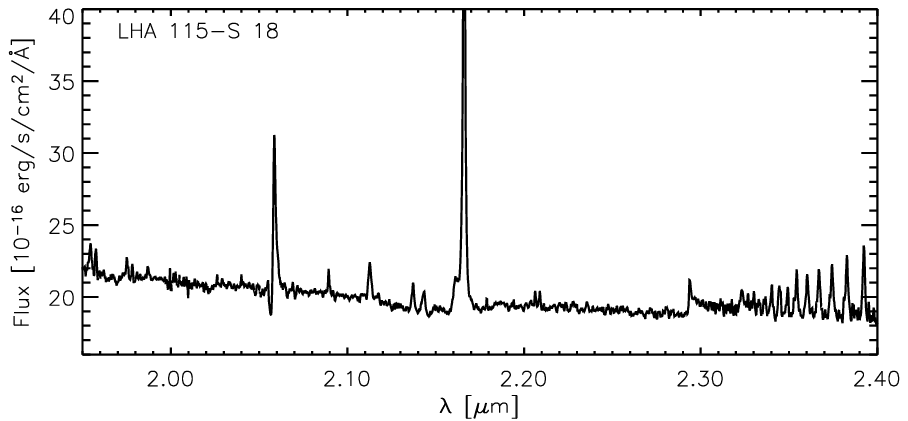} &
\includegraphics[width=80mm]{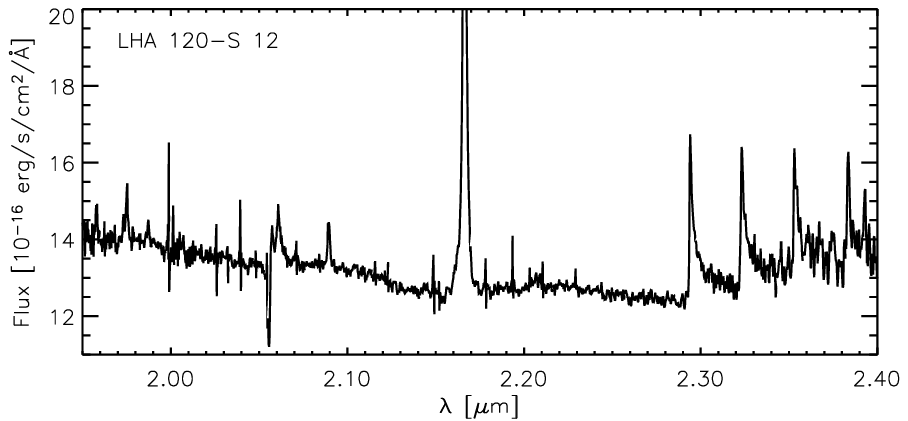} \\
\includegraphics[width=80mm]{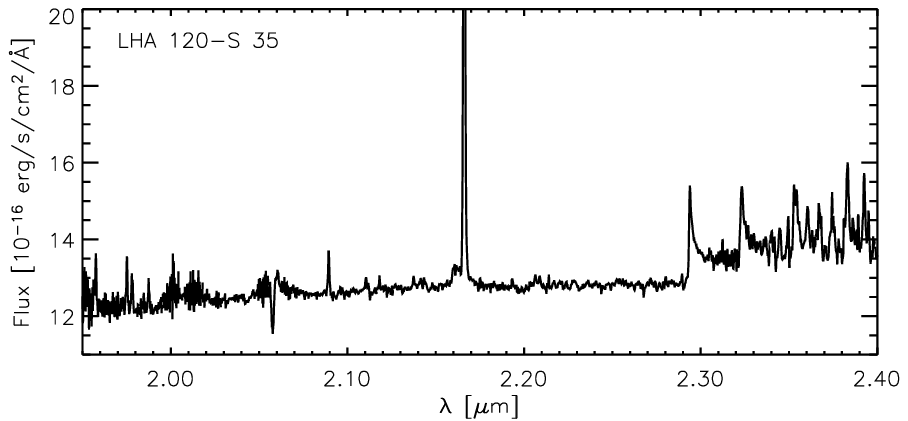} &
\includegraphics[width=80mm]{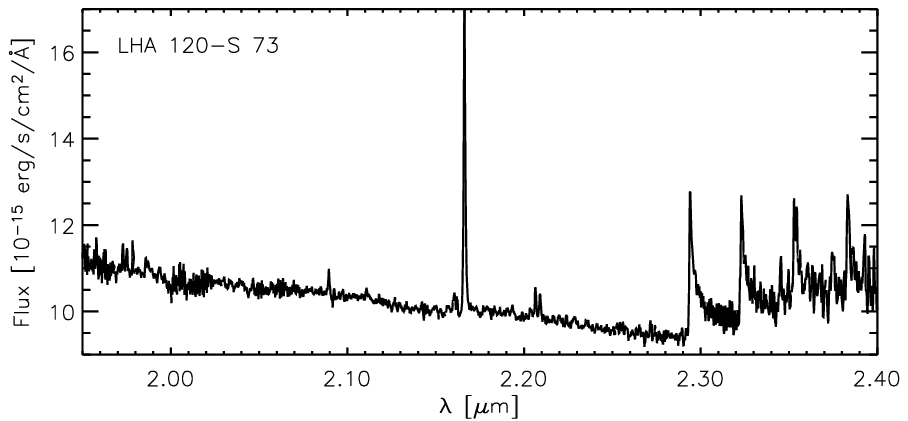} \\
\includegraphics[width=80mm]{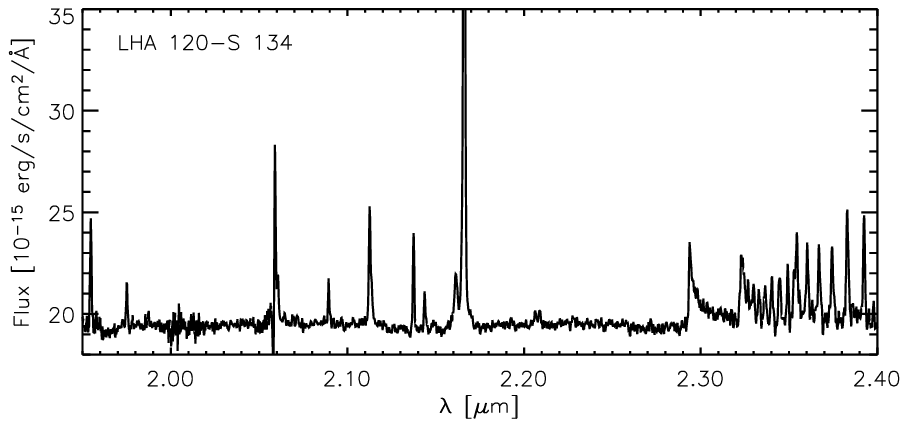} &
\includegraphics[width=80mm]{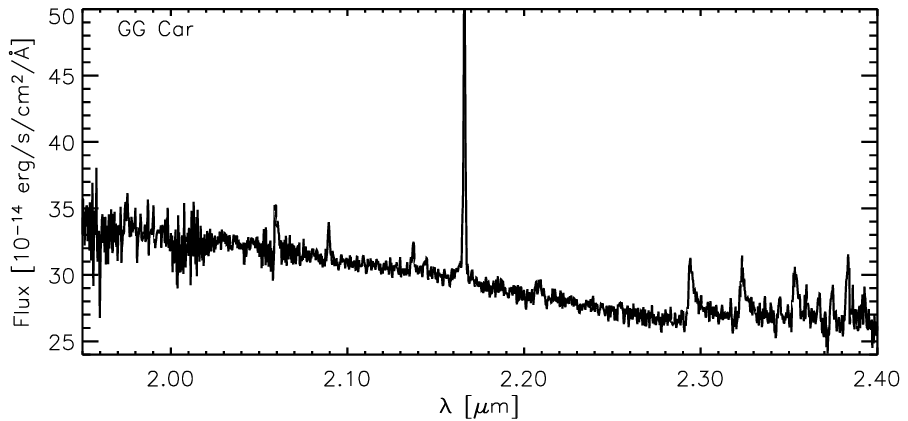} \\
\includegraphics[width=80mm]{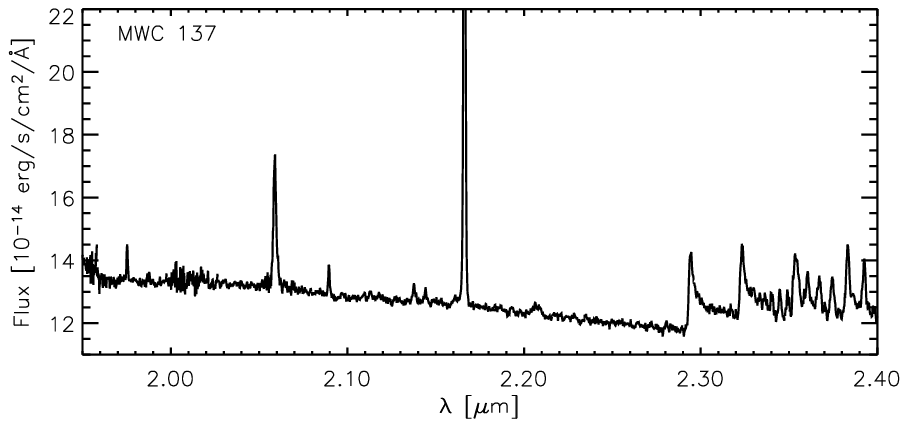} &
\includegraphics[width=80mm]{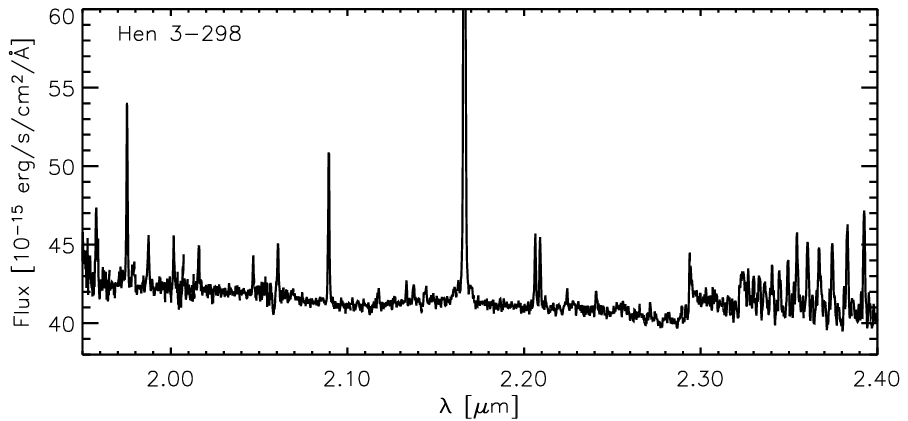}
\end{tabular}
\caption{Flux-calibrated SINFONI $K$-band spectra of B[e]SGs with definite CO band head emission}
\label{FULLspec1}
\end{figure*}

\begin{figure*}
\centering
\begin{tabular}{cc}
\includegraphics[width=80mm]{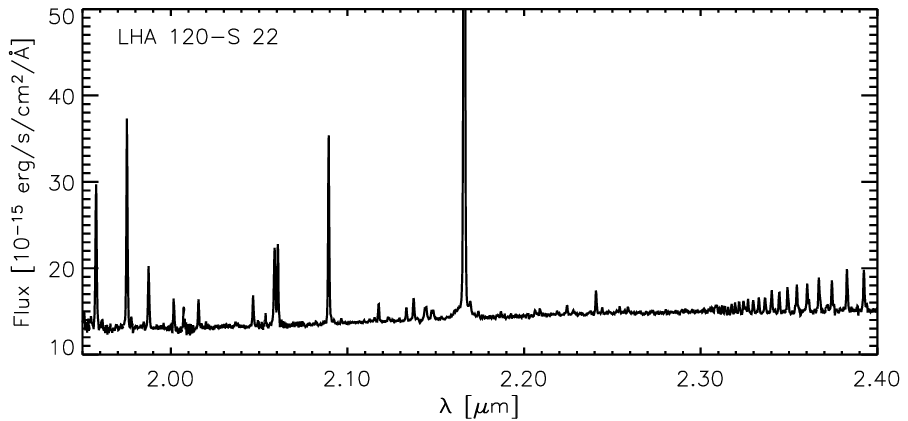}&
\includegraphics[width=80mm]{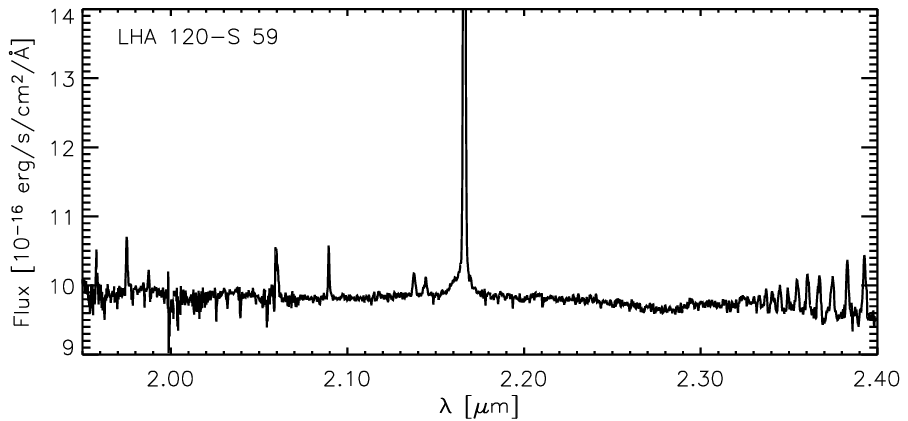} \\
\includegraphics[width=80mm]{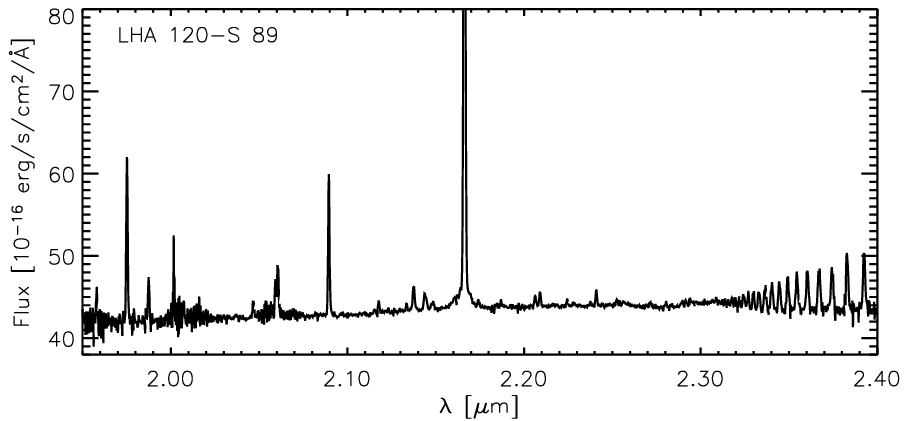} &
\includegraphics[width=80mm]{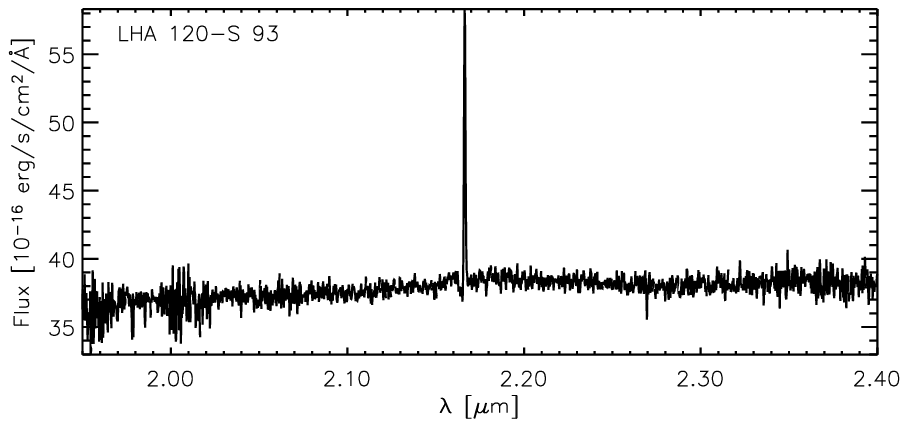} \\
\includegraphics[width=80mm]{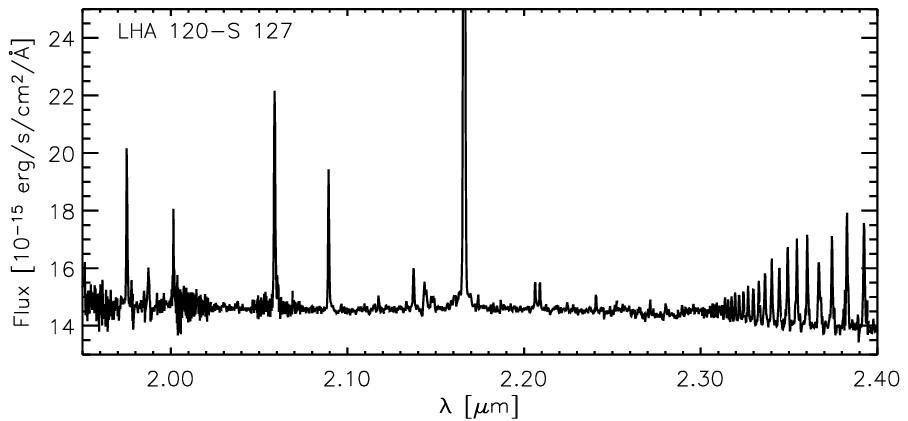} &
\includegraphics[width=80mm]{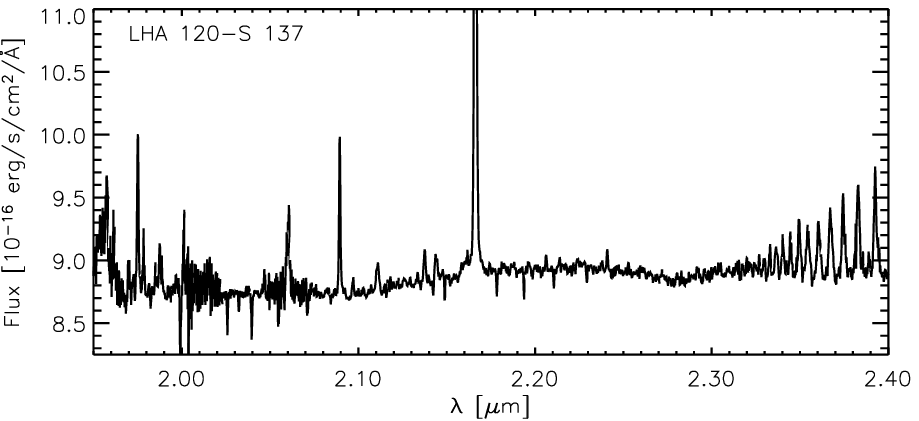} \\
\end{tabular}
\caption{Flux-calibrated SINFONI $K$-band spectra of B[e]SGs with absent or uncertain CO band head emission.}
\label{FULLspec2}
\end{figure*}

\begin{table}
\caption{Observation Summary. 2MASS $K_{s}$-band magnitudes derived from 
the flux calibrated spectrum.}             
\label{tbl-2}      
\centering          
\begin{tabular}{lccccc} 
\hline\hline 
Object & Date & Time & t$_{\rm{EXP}}$ & Final & $K_{s}$ \\
 & & UT & (s) & S/N  & (mag) \\
\hline 
LHA 115-S 6    &   2011-10-05 & 01:48:43  & 150  & 200  & 11.18  \\
------------   &   2011-10-06 & 04:09:34  & 150  & --   & --    \\
LHA 115-S 18   &   2009-10-12 & 01:56:49  & 300  & 100  & 10.90  \\
------------   &   2009-10-16 & 02:27:33  & 300  & --   & --    \\
LHA 115-S 52   &   2011-10-06 & 04:55:07  & 30   & 150  & 8.77   \\
LHA 115-S 65   &   2011-10-06 & 05:26:58  & 60   & 200  & 9.63   \\
HD 269723      &   2011-12-23 & 00:31:35  & 10   & 150  & 7.69   \\
HD 269953      &   2012-01-21 & 03:46:10  & 10   & 150  & 8.15   \\
LHA 120-S 12   &   2009-10-14 & 07:55:34  & 150  & 150  & 11.36  \\
LHA 120-S 22   &   2011-12-23 & 00:31:35  & 20   & 150  & 8.73   \\
LHA 120-S 35   &   2012-02-16 & 00:18:47  & 150  & 200  & 11.33  \\
------------   &   2012-02-16 & 02:43:18  & 150  & --   & --    \\
LHA 120-S 59   &   2011-11-20 & 07:38:41  & 150  & 250  & 11.60  \\
------------   &   2012-01-10 & 05:13:49  & 150  & --   & --    \\
------------   &   2012-02-04 & 01:10:49  & 150  & --   & --    \\
LHA 120-S 73   &   2009-11-09 & 07:09:21  & 100  & 200  & 9.10   \\
LHA 120-S 89   &   2012-02-01 & 00:58:53  & 60   & 300  & 10.03  \\
LHA 120-S 93   &   2012-01-26 & 02:40:25  & 150  & 100  & 10.19  \\
LHA 120-S 96   &   2011-10-11 & 08:35:59  & 20   & 150  & 9.13  \\
LHA 120-S 116  &   2012-02-04 & 02:05:39  & 60   & 150  & 9.74   \\
LHA 120-S 124  &   2012-02-04 & 02:28:10  & 60   & 100  & 9.60   \\
LHA 120-S 127  &   2012-02-05 & 01:21:52  & 30   & 200  & 8.72   \\
LHA 120-S 128  &   2012-02-05 & 01:33:10  & 30   & 150  & 9.28   \\
LHA 120-S 134  &   2009-11-10 & 08:25:07  & 30   & 150  & 8.43   \\
LHA 120-S 137  &   2012-01-21 & 03:46:10  & 150  & 300  & 11.78  \\
------------   &   2012-01-21 & 04:05:55  & 150  & --   & --    \\
LHA 120-S 155  &   2012-01-26 & 01:47:05  & 60   & 100  & 7.09   \\
MWC 137        &   2011-10-05 & 08:46:49  & 2    & 200  & 6.68   \\
GG Car         &   2012-01-09 & 05:36:17  & 0.83 & 60   & 5.55   \\
Hen 3-298      &   2012-01-09 & 05:55:56  & 1    & 200  & 5.56   \\
WRAY 15-751    &   2012-01-08 & 08:00:00  & 1    & 200  & 4.16   \\
\hline
\end{tabular}
\end{table}

\subsection{Sample selection}

We obtained $K$-band spectroscopic data for a near-infrared study of the 
molecular circumstellar material and possible evolutionary links between various phases of 
massive star evolution.  The target stars in this sample were chosen to highlight 
possible physical stages of massive star evolution that may be connected, with an emphasis 
on the group of particularly perplexing B[e]SGs.  The sample contains 25 massive evolved 
stars in the Galaxy and the Magellanic Clouds, including 16 B[e]SGs, 6 LBVs, 2 YHGs, and 
1 Peculiar Oe star.  Each target star is listed in Table~\ref{tbl-1} together with various 
properties and stellar parameters derived from the literature.  
In the case of LHA 115-S 6, we consider it a B[e]SG \citep{Zickgraf87}, although it has 
shown characteristics typical of an LBV \citep{Zickgraf96}.
Included in this study are all of the known Magellanic Cloud B[e]SGs studied by \citet{Zickgraf86}, 
with the exception of LHA 120-S 111 
in the Large Magellanic Cloud (LMC) and LHA 115-S 23 in the Small Magellanic Cloud (SMC).
These two stars were excluded because of their faint magnitudes.  The chosen targets represent 
not only distinct evolutionary stages, but also a variety of initial masses, luminosities, 
and effective temperatures.  Many of the objects have $K$-band spectra featured in the works of 
\citet{McGregor88a}, \citet{McGregor88b}, and \citet{McGregor89}, including the first 
detection of CO emission in an evolved massive star \citep{McGregor88a}, although for several objects 
this is the first observation of this wavelength region.  

\begin{figure*}[!ht]
\centering
\begin{tabular}{cc}
\includegraphics[width=80mm]{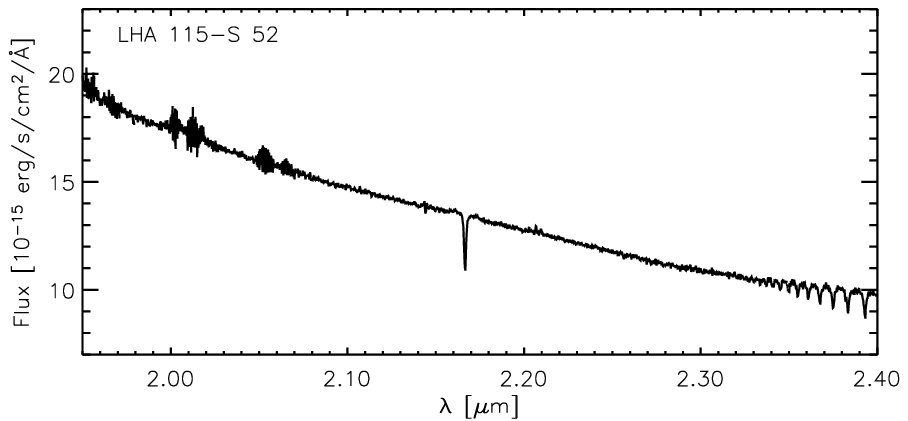}&
\includegraphics[width=80mm]{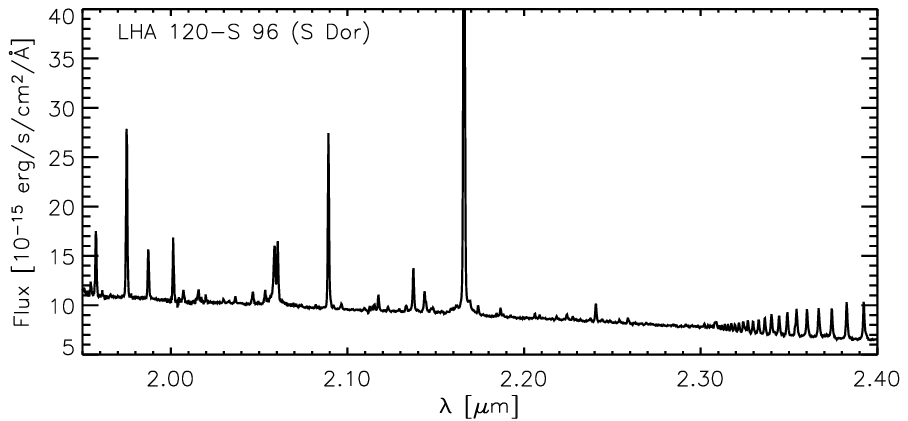} \\
\includegraphics[width=80mm]{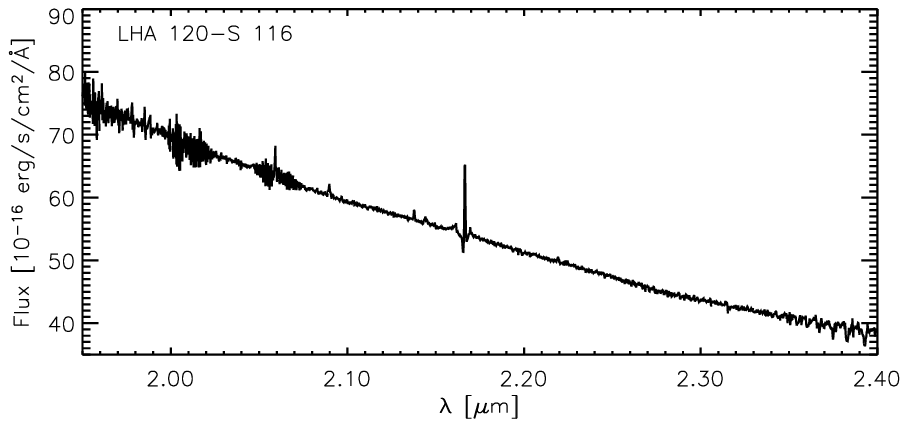} & 
\includegraphics[width=80mm]{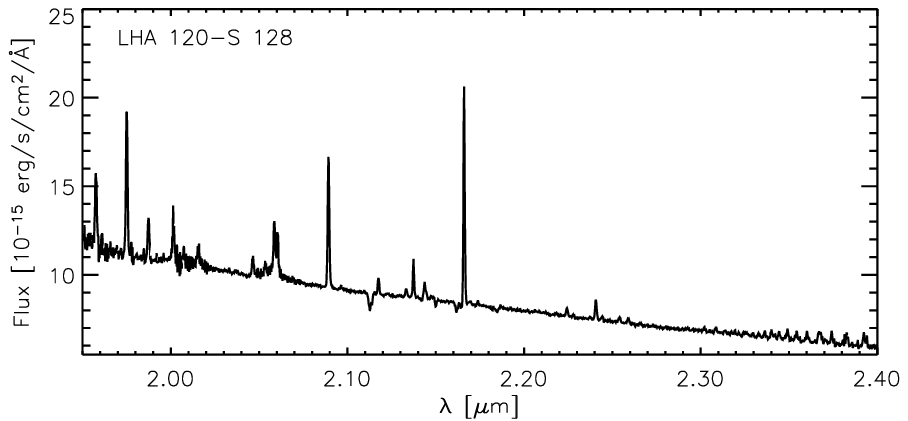} \\
\includegraphics[width=80mm]{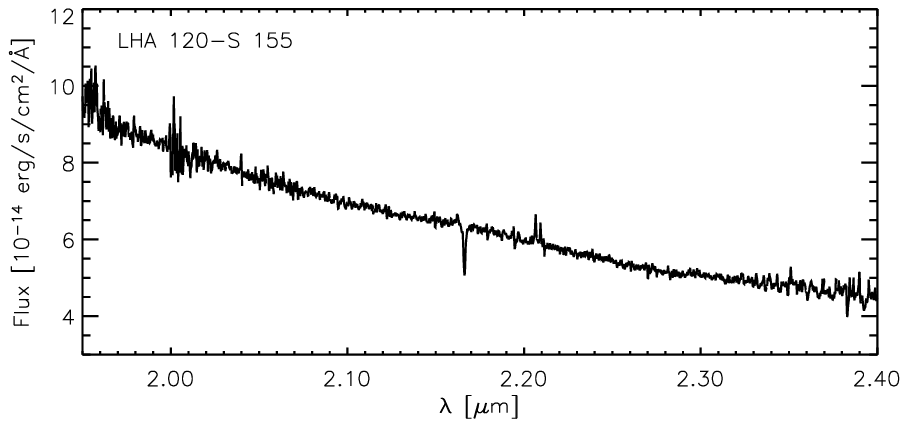}&
\includegraphics[width=80mm]{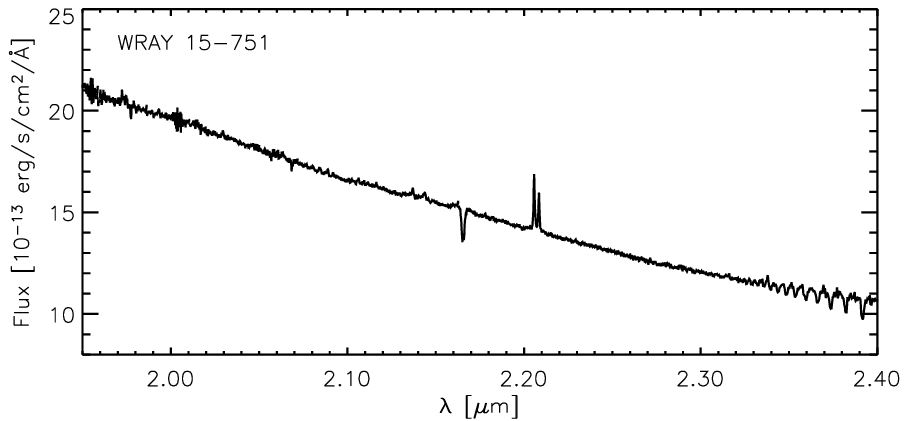}
\end{tabular}
\caption{Flux-calibrated SINFONI $K$-band spectra of LBVs.}
\label{FULLspec4}
\end{figure*}

\subsection{Data acquisition and reduction}\label{red}

For each sample target, a medium resolution (R=4500) $K$-band spectrum was obtained
with the Spectrograph for INtegral Field Observation in the Near-Infrared 
\citep[SINFONI;][]{Eisenhauer03, Bonnet04}, on the ESO VLT UT4 8~m telescope.
The observations were obtained from 2009-2012 (individual dates of observation and exposure times
are listed in Table~\ref{tbl-2}).  The spectrograph was set to an 8$\times$8 arcsec$^{2}$ field of view 
(corresponding to a 250 mas spatial resolution).  The observations were taken in an ABBA 
nod pattern for optimal sky subtraction.  A B-type standard star was observed immediately 
following each target star and at a similar airmass value to perform both telluric 
correction and flux calibration.

\begin{figure}[!ht]
\centering
\begin{tabular}{c}
\includegraphics[width=80mm]{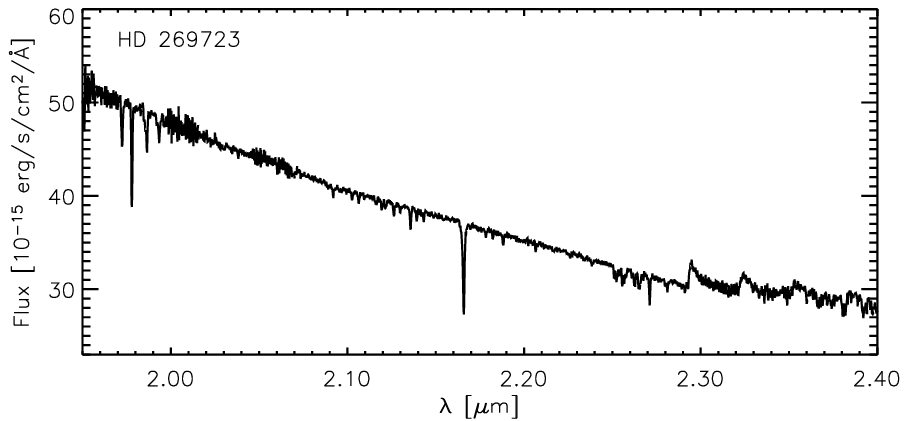}\\
\includegraphics[width=80mm]{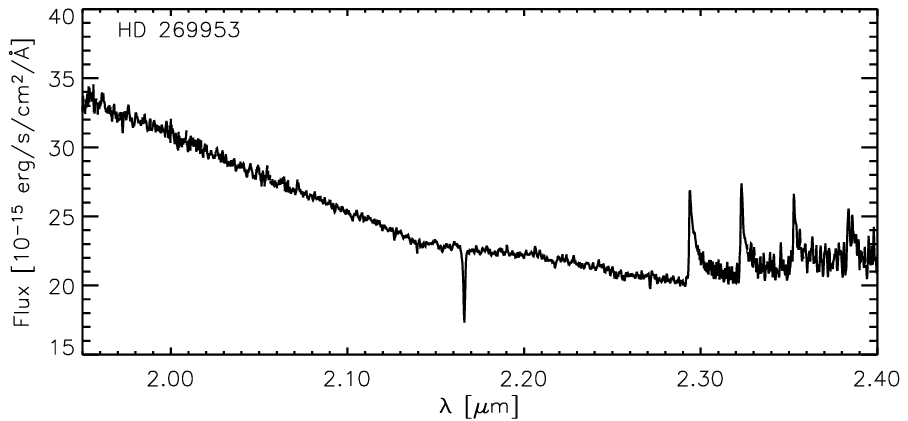}  \\
\includegraphics[width=80mm]{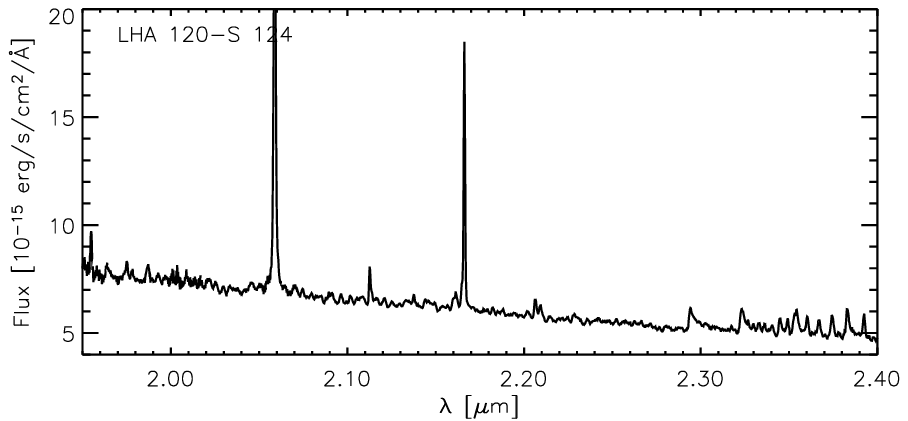}  \\
\end{tabular}
\caption{Flux-calibrated SINFONI $K$-band spectra of the YHGs HD 269953 and HD 269723 and the Peculiar Oe star LHA 120-S 124.}
\label{FULLspec3}
\end{figure}

The raw data for both the target and standard stars were reduced using the SINFONI 
pipeline (version 2.2.9).  The observations were treated for bad pixels, flat fields, 
and distortions, and then wavelength calibrated.  In the infrared, there can be a large amount 
of telluric lines that can be difficult to remove.  We chose to use main sequence B-type standards, since these stars 
have few spectral features in the infrared wavelength region, typically lines of hydrogen and helium.  Using these 
standard spectra, we created telluric templates and corrected each target spectrum 
using the IRAF\footnote{IRAF is distributed by the National Optical Astronomy Observatory, which is operated by the Association of Universities for Research in Astronomy (AURA) under cooperative agreement with the National Science Foundation.} task \textit{telluric}.  Some areas of telluric contamination were particularly difficult 
to correct, leaving remnant traces in these specific wavelength regions. To flux calibrate each target spectrum, the 
corresponding standard star spectrum was scaled with an appropriate Kurucz flux model 
\citep{Kurucz93} to its Two Micron All-Sky Survey (2MASS) \citep{Skrutskie06} $K_{s}$-band magnitude.

The resultant spectra were then corrected (using the values listed in Table~\ref{tbl-1}) 
for heliocentric and systemic velocities and dereddened 
with the appropriate corresponding E(B-V) values, according to the interstellar extinction 
relation of \citet{Howarth83} assuming R=3.1.  The final spectra have signal-to-noise (S/N) 
ratios ranging from 60-300 (listed in Table~\ref{tbl-2}).  Figs.~\ref{FULLspec1}-\ref{FULLspec3} 
show the full spectrum for each of our targets. The $K$-band of the SINFONI spectrograph
has a wavelength range of 1.95-2.45 $\mu$m, containing the Brackett hydrogen
recombination line (Br$\gamma$) at 2.167 $\mu$m,  the upper levels of the Pfund series,
and various helium and metal lines.  Additionally, the first overtone band heads for the CO molecule 
are located within this region.

\begin{figure}[!ht]
\centering
\includegraphics[width=80mm]{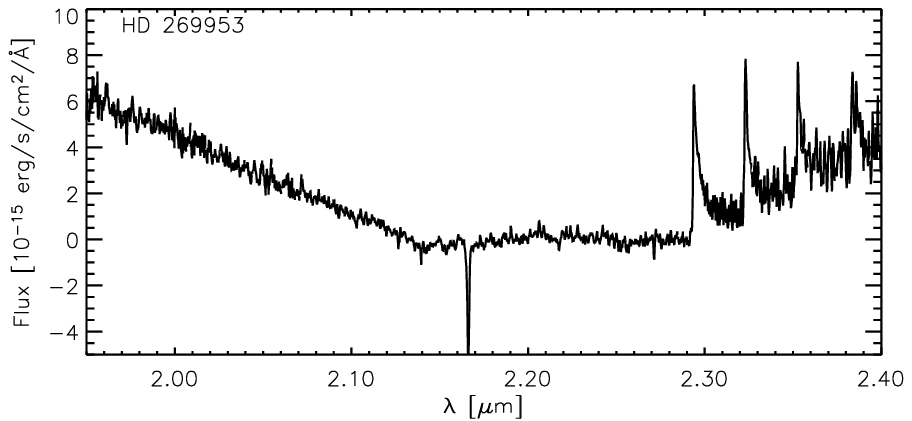}
\includegraphics[width=80mm]{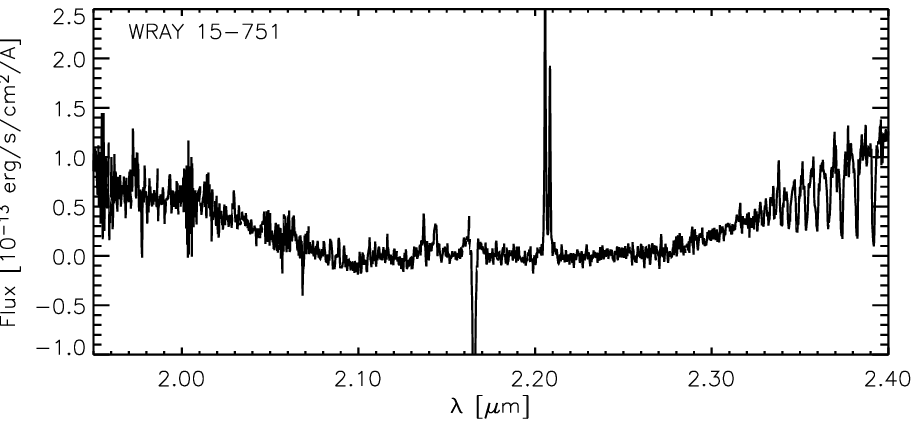}
\caption{Examples of the group of objects with cool supergiant type spectra.  Flux-calibrated, continuum subtracted SINFONI $K$-band spectra of the YHG HD 269953 and the LBV WRAY 15-751.}
\label{CONTspec2}
\end{figure}

\onlfig{
\begin{figure*}[!ht]
\centering
\begin{tabular}{cc}
\includegraphics[width=80mm]{normHD269953-1spc.eps} &
\includegraphics[width=80mm]{normWRAYspc.eps} \\
\includegraphics[width=80mm]{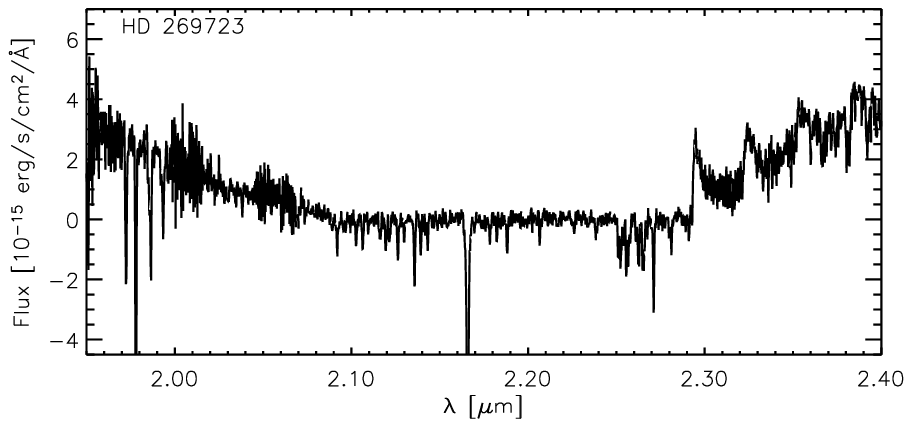} &
\includegraphics[width=80mm]{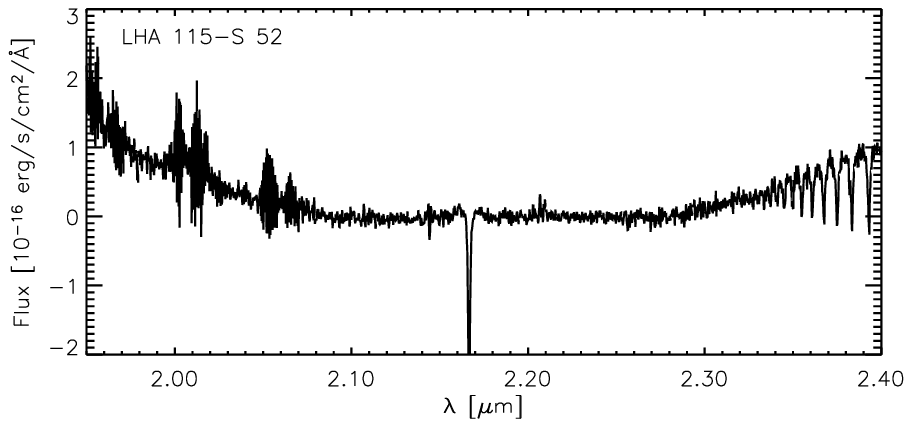} \\
\includegraphics[width=80mm]{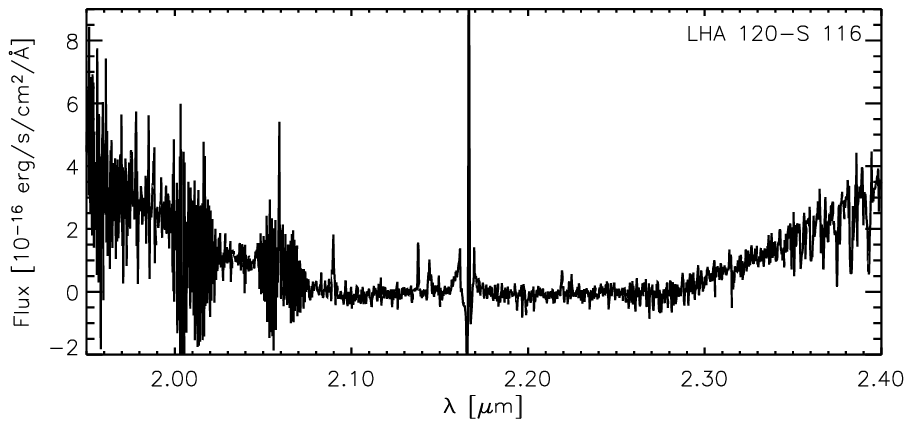} &
\includegraphics[width=80mm]{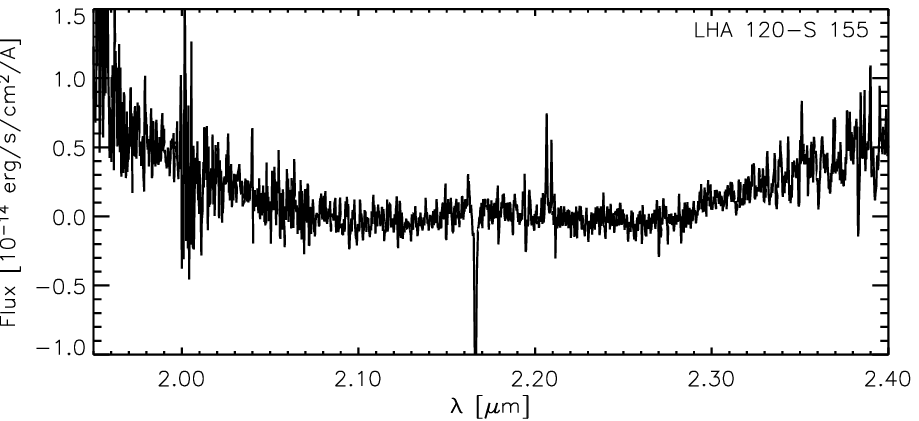} 
\end{tabular}
\caption{Objects with cool supergiant type spectra.  Flux-calibrated, continuum subtracted SINFONI $K$-band spectra of the YHGs HD 269953 and HD 269723 and the LBVs WRAY 15-751, LHA 115-S 52, LHA 120-S 116, LHA 120-S 155.}
\label{CONTspec3}
\end{figure*}}

\begin{figure}[!ht]
\centering
\includegraphics[width=80mm]{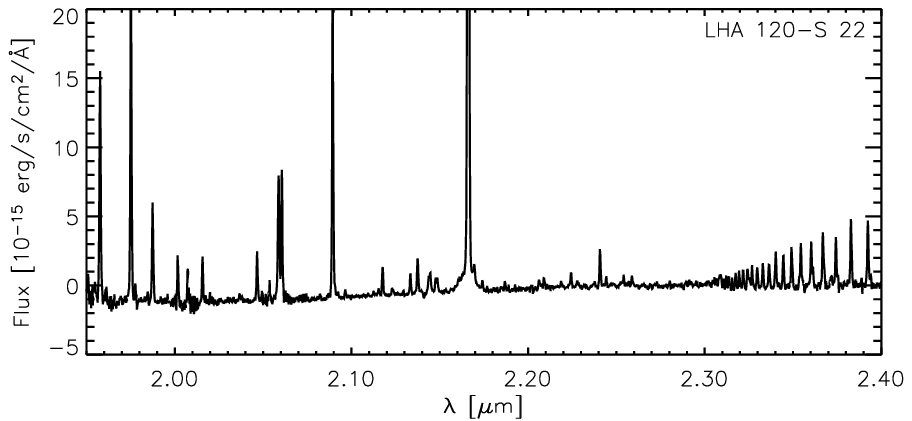}
\includegraphics[width=80mm]{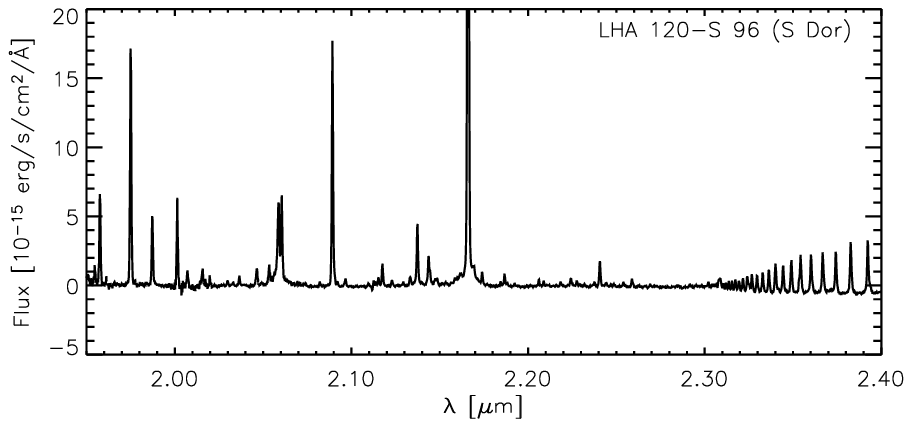}
\caption{Examples of the group of objects with hot supergiant type spectra.  Flux-calibrated, continuum subtracted SINFONI $K$-band spectra of the B[e]SG LHA 120-S 22 and the 
LBV LHA 120-S 96 (S Dor).}
\label{CONTspec1}
\end{figure}

\begin{figure}
\centering
\includegraphics[width=80mm]{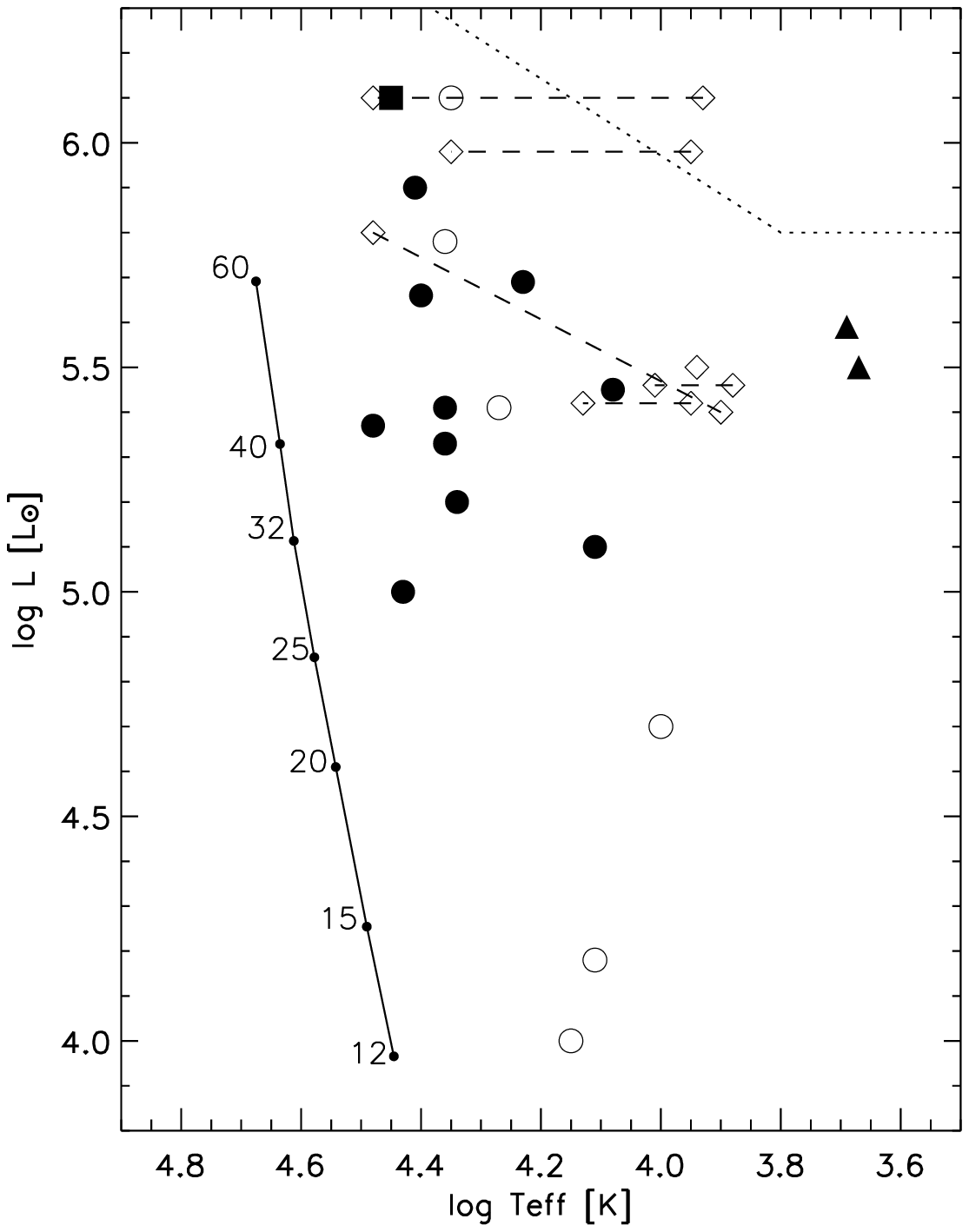}
\caption{Hertzsprung-Russell Diagram (HRD) showing the location of each of our target stars.  
The start of the main sequence is indicated by the solid line, with specific initial masses 
from the evolutionary models of \citet{Ekstrom}.  Filled (open) symbols are stars with definite 
(absent or uncertain) CO band head emission. Based on the object types in Table~\ref{tbl-1}, 
circles are B[e]SGs, diamonds represent LBVs, triangles show YHGs, and the square is the Peculiar Oe star.  
Dashed lines between symbols indicate LBV variable parameters.
Also plotted is the Humphreys-Davidson limit \citep[dotted line,][]{HD94}.}
\label{HRplot2}
\end{figure}

\subsection{$K$-band Magnitudes}\label{Kmag}

From the flux calibrated final spectra, we computed the 2MASS $K_{s}$ (K short) magnitude at the time of observation
for each target.  The calculation utilized the filter function and calibration constant (zero point flux) given by \citet{Cutri}, as
described in \citet{Liermann09}.  
These values are listed in the last column
of Table~\ref{tbl-2}. The process of obtaining flux calibrated spectra contains uncertainties, 
such as stability of the magnitude and the effective temperature assumed for the standard star.  
Also, the telluric correction process can influence the flux level.  Comparison between the 2MASS
$K_{s}$-band measurements and these newly determined values indicate that in most cases the
two values are similar with differences typically below $\sim$0.5 mag, likely due to the aforementioned uncertainties.
In several cases though, the change in magnitude appears significant.  The LBVs LHA 115-S 52, Wray 15-751, and LHA 120-S 155
all show brighter magnitudes, with the latter two significantly increased by 2.6 and 3.5 magnitudes, respectively.  The LBV LHA 120-S 96 (S Dor) 
shows a slight decrease in brightness.  The B[e]SG LHA 120-S 12 is the only non-LBV that shows notable magnitude change, with
a decrease in brightness of approximately one magnitude.

\subsection{Continua}\label{continua}

To properly model the pure CO emission, we subtracted a 
continuum from the spectrum (described in Section~\ref{models}, fit using the portion
of the spectrum from $\sim 2.17-2.25~\mu$m).  While 
the area of interest for the modeling procedure is only a small section of 
the spectrum, the ``normalization'' of each spectra produced two distinct 
groups (i) kinked and (ii) generally flat spectra, based solely on 
the shape of their continua. 

Stars within the first group display a shape of the continuum featuring two 
bends or kinks, one at $\sim2.1$ $\mu$m and another at $\sim2.3$ $\mu$m.
Such a behavior is already noticeable for one object, the YHG HD 269953
(see Fig.~\ref{CONTspec2}) upon visual inspection of its flux calibrated spectrum without 
normalization. Because of the broad range of flux in Figs.~\ref{FULLspec1}-\ref{FULLspec3}, 
other objects that fall into this category became clear only after normalization.  The remaining 
members of this group are the 
second YHG star, HD 269723, and the LBVs LHA 115-S 52, LHA 120-S 116, 
LHA 120-S 155, and WRAY 15-751.  The spectrum in all these stars slopes 
upward blue-ward of the first bend and slopes upward again red-ward of 
the second.  The slope in the red portion of the spectrum may be due to 
the presence of dust in the circumstellar environment, while the blue ward 
increasing slope indicates a 
cool stellar photospheric spectrum in all these objects (see Sect~\ref{LBVstate}).
The location of the bends depend on the physical conditions of the dust, 
the strength of the stellar wind, and the temperature of the star. 
To highlight this kinked shape, Fig.~\ref{CONTspec2} shows continuum 
subtracted spectra of two examples, the YHG HD 269953 and the LBV WRAY 15-751.
Fig.~\ref{CONTspec3} (available online only) contains a plot showing
all six of the objects of this group.

In the other group of objects, the continuum throughout the entire $K$-band
wavelength range is generally flat.  Several target spectra do have an
increasing slope in the red most portion of the spectrum in the location
of the CO band heads and/or Pfund lines, due to the presence of a hot dust
continuum.  The objects in this group include all of the B[e]SGs, the
Peculiar Oe star LHA 120-S 124, and the LBVs LHA 120-S 96 (S~Dor) and
LHA 120-S 128.  Two examples of this second group of
objects are shown in Fig.~\ref{CONTspec1}, the B[e]SG LHA 120-S 22
and the LBV S Dor.  As the figure shows, visually the B[e]SGs and these
LBVs are indistinguishable.

\subsection{Spectral Composition}

\onllongtab{
\begin{landscape}
\begin{longtable}{lcccccccccccccccc}    
\caption{\label{tbl-5} Line identification and equivalent width measurements for B[e]SGs. Measurements are in units of \AA~and for emission lines unless indicated by an A, indicating the line is in absorption.  Bl indicates a blended line, while SL indicates shell line.}  \\    
\hline\hline 
\noalign{\smallskip}
Line & S6 & S18 & S65 & S12  & S22 & S35  & S59 & S73 & S89 & S93 & S127 & S134  & S137 & MWC 137 & Hen 3-298 & GG Car \\
\noalign{\smallskip}
\hline
\endfirsthead
\caption{continued.}\\
\hline\hline
\noalign{\smallskip}
Line & S6 & S18 & S65 & S12  & S22 & S35  & S59 & S73 & S89 & S93 & S127 & S134  & S137 & MWC 137 & Hen 3-298 & GG Car \\
\noalign{\smallskip}
\hline
\endhead
\hline
\tablefoottext{a}{P Cygni profile}
\endfoot
\ion{He}{i}~$\lambda1.949$	&  ...	&  ...		&  1.95	 &  ...	&  ...		&   ...   &  ...	&  ...	 & ...	  	&  ...	 &   ...	&  ...		& ... 		& ... 		& ... 		&  ... \\
?~$\lambda1.955$	&  ...	&  1.78	&   ...	 &  ...	&  ...		&  ...    &  ...	&  ...	 & ...	  	&  ...	 &    ...	&  2.93	& ...		& ... 		& ... 		&  ... \\
\ion{Fe}{ii}~$\lambda1.957$          		&  1.34	&  1.11	&   3.20	 &  0.78	&   12.59	& 1.37 &  0.59 	&  ...	 & 0.86  	&  ...	 &    ...	& ...		& 0.99	& 0.34	& 1.83	&  ... \\
?~$\lambda1.963$	&  0.45	&   ...	&   ...	 &  ... 	&   ...	&  ...    &  ...	& ...	 &  ...	&  ...	 &    ...	& ...		& ...		& ... 		& ... 		& ... \\
{[\ion{Fe}{ii}]}~$\lambda1.968$	&  0.31	&   ...	&   ...	 & ...		&   ...	&   ...   &  ...	& ...	 &   ...	&  ...	 &    ...	& ...		& ...		& ... 		& ... 		& ... \\
\ion{Fe}{ii}~$\lambda1.976$&  0.45	&  0.66	&   5.52	 &  1.88	&   14.70	&  ...    &  ...	&  ...	 &   5.03	&  ...	 &    2.50	& 1.06	& 1.69	& 0.72	& 2.19	&  ... \\
?~$\lambda1.977$	&  ...		&  ...		&   0.96	 &  ...	&   ...	&  ...	   & 0.97	&  ...	 &   ...	&  ...	 &   ...	&  ...		& ...		& ... 		& ... 		&  ... \\
\ion{Ca}{i}~$\lambda1.978$ 			&  1.79	&  0.43	&   0.37   	 &  ...	&   ...	&  ...	   &  0.93	&  ...	 &   ...	&  ...	 &   ...	&  ...		& 0.38	& ... 		& ... 		& ...  \\
?~$\lambda1.985$	&  ...		&  ...		&   1.48	 &  ...	&   ...	&  ...	   &  ...	&  ...	 &   ...	&  ...	 &    ...	& ...		& ...		& ... 		& ... 		&  ... \\
\ion{Ca}{i}~$\lambda1.987$      &  0.86 	&  0.54	&   1.47	 &  0.70	&   4.56	&  ...	   &  0.26	&  ...	 &   2.11	&  ...	 &   0.88	& ...		& 0.38	& 0.21 	& 0.93	&  ... \\
?~$\lambda2.000$	&  1.42	&  ...		&   ...	 &  SL	&   ...	&  ...	   &  ...	&  ...	 &   ...	&  ...	 &    ...	& ...		& 0.44A	& ... 		& ... 		&  ... \\
?~$\lambda2.002$         &  1.30	&  0.51	&   1.09	 &  0.49	&   2.17	&  ...	   &  ...	&  ...	 &   1.47	&  ...	 &   0.96	& ...		& ...		& ... 		& 0.64	&  ... \\
{[\ion{Fe}{ii}]}~$\lambda2.008$ 	&   ... &  ... 	&   ...	 &  ...	&   1.63	&  ...	   &  ...	&  ...	 &   ...	&  ...	 &    ...	& ...		& ...		& ... 		& 0.35	& ...  \\
{[\ion{Fe}{ii}]}~$\lambda2.016$ 	&   ...	&  ... 	&   1.18	 &  ...	&   1.80	&  ...	   &  ...	&  ...	 &   ...	&  ...	 &    ...	& ...		& ...		& ... 		& 0.54	& ...  \\
?~$\lambda2.027$              &  0.45 &  ... 	&   ...	 &  SL	&   ...	&  ...	   & 0.19A&  ... &   ...		&  ...	 &    ...	& ...		& 0.21A	& ... 		& ... 		&  ... \\
?~$\lambda2.030$         	&   ...	&  ... 	&   ...	 &  SL	&   ...	&  ...	   & 0.13A&  ... &   ...		&  ...	 &    ...	& ...		& 0.11A	& ... 		& ... 		&  ... \\
H$_{2}~\lambda2.034$           &   0.30&  ... 	&   ...	 &  ...	&   ...	&  ...	   &  ...	 &  ... &   ...	&  ...	 &    ...	& ...		& ...		& ... 		& ... 		& ...  \\
?~$\lambda2.041$        &   0.68&  ... 	&   ...	 &  SL	&   ...	&  ...	   & 0.33A& ...	  &   ...	&  ...	 &    ...	& ...		& 0.28A	& ... 		& ... 		&  ... \\
{[\ion{Fe}{ii}]}~$\lambda2.046$	 &   ...	&  ... 	&   0.88	 &  SL	&   2.06	&  ...	   &  ...	 & ...	  &   0.50	&  ...	 &    ...	& ...		& ...		& ... 		& 0.46	& ...  \\
{[\ion{Fe}{ii}]}~$\lambda2.054$ &   0.95&  ... 	&   ...	&  2.89A&   ...		&  ...	   & 0.24A & ...  &   ...	&  ...	 &  ...	& ...		& 0.26A	& ... 		& ... 		& ...  \\
\ion{He}{i}~$\lambda2.058$     	&   3.53&  6.88\tablefootmark{a}&  0.43  & 0.62 &  6.40 & 1.54A  & 0.96\tablefootmark{a}  & ...  &  1.33 &  ... &  4.76 & 2.53\tablefootmark{a}& 1.13\tablefootmark{a} & 6.34 & 0.39A & 2.12\tablefootmark{a} \\
\ion{Fe}{ii}~$\lambda2.061$      &   2.52&  Bl &   1.89	 &  2.17	&   5.50	&  ...	   & Bl  &  ... &  1.81 &  ...	 & 0.44  & 1.21& Bl & Bl & 0.79 & Bl \\
?H$_{2}~\lambda2.074$           &   0.46&  ... 	&   ...	 &  SL	&   ...	&  ...	   & ...	  & ...  &  ...	 &  ... &   ...	& ...		& ...		& ... 		& ... 		& ...  \\
\ion{Fe}{ii}~$\lambda2.091$     &   1.87&  0.61	&   4.83	 &  1.43	&   11.48	&  0.90& 0.75	  &  0.48 & 3.84  &  ... &   2.44	& 1.08	& 1.45	& 0.80	& 2.14	& 1.26 \\
?~$\lambda2.105$               			&   ...	&  ... 	&   0.58 	 &  ...	&   ...	&  ...	   &  ...	 &  ... &   ...	&  ...	 &    ...	& ...		& ...		& ... 		& ... 		& ...  \\
\ion{He}{i}~$\lambda2.112/3$    &   ...	&  2.42	&   ...	 &  ...	&   ...	&  ...	   &  ...	 &  ... &   ...	&  ...	 &    ...	& 4.08	& 0.47	& ... 		& ... 		&  ... \\
\ion{He}{i}~$\lambda2.114$     	&   ...	&  ... 	&   ...	 &  ...	&   ...	&  ...	   &  ...	 & ...  &   ...	&  ...	 &    ...	& ...		& ...		& ... 		& ... 		&  ... \\
?\ion{N}{iii}/\ion{C}{iii}~$\lambda2.116$ 	&   ...	&  ... 	&   0.45 	 &  SL	&   ...	&  ...	   &  ...	 &  ... &   ...	&  ...	 &    ...	& ...		& ...		& ... 		& ... 		& ...  \\
?{[\ion{Fe}{ii}]}/\ion{Al}{i}~$\lambda2.118$ &   ...	&  ... 	&   0.46 	 &  ...	&   1.20	&  ...	   &  ...	 &  ... &   0.33	&  ...	 &   0.31	& ...		& ...		& ... 		& 0.30 	& ...  \\
?H$_{2}$/\ion{Fe}{ii}~$\lambda2.124$ 	&   ...	&  ... 	&   ...	 &  SL	&   ...	&  ...	   &  ...	 &  ... &   ...	&  ...	 &    ...	& ...		& ...		& ... 		& ... 		& ...  \\
{[\ion{Fe}{ii}]}~$\lambda2.133$ 	&   ...	&  ... 	&   0.77	 &  ...	&   0.83	&  ...	   &  ...	 &  ... &   0.20	&  ...	 &    ...	& ...		& ...		& ... 		& 0.21 	& ...  \\
\ion{Mg}{ii}~$\lambda2.137$      	&   0.97	&  1.90	&   0.43	 &  ...	&   1.62	&  ...	   &  0.58	 &  ... &   0.89	&  ...	 &   0.90	& 2.28	& 0.34	& 0.76	& 0.35	&  1.28 \\
\ion{Mg}{ii}~$\lambda2.144$      &   0.59	&  1.91	&   0.85 	 &  ...	&   1.65	&  ...	   &  0.44	 &  ... &   0.84	&  ...	 &   1.00	& 1.33	& 0.49	& 0.48	& 0.40	& 0.70  \\
?{[\ion{Fe}{ii}]}/\ion{He}{i}~$\lambda2.148$ &  ... 	&  ... 	&   0.44	 &  SL	&   1.15	&  ...	   &  ...	 &  ... &   0.32	&  ...	 &   0.56	& ...		& ...		& ... 		& ... 		& ...  \\
\ion{He}{i}~$\lambda2.150$      &   0.55	&  ... 	&   ...	 &  ...	&   ...	&  ...	   &  ...	 &   ...&   ...	&  ...	 &    ...	& ...		& ...		& ... 		& ... 		& ...  \\
H$_{2}~\lambda2.154$           &   0.22	&  ... 	&   ...	 &  SL	&   ...	&  ...	   &  ...	 &  ... &   ...	&  ...	 &    ...	& ...		& ...		& ... 		& ... 		& ...  \\
\ion{He}{i}~$\lambda2.161$      & Bl  & 6.72 &  Bl  &  4.32  &  Bl & ...  &  ... &  0.29 &  Bl &  ...	 & 0.44 & 3.44 & Bl & 0.74 &  Bl &  Bl \\
\ion{H}{i}~Br~$\gamma$~$\lambda2.166$ &   30.23	& 32.95 &  46.13  & 38.02 & 75.67  & 21.17  & 19.25  &  8.59 & 31.97  & 4.58 &  42.56 & 26.41 & 16.13 & 30.28 & 23.68 & 10.41  \\
\ion{He}{i}~$\lambda2.166$      &  Bl &  Bl    &  Bl  & Bl       &  Bl & ...  &  ... & Bl  & Bl  &  ... &  Bl & Bl & Bl & Bl & Bl & Bl \\
?~$\lambda2.169$               &   ...	&  ... 	&   ...	 &  ...	&  Bl   &   ...	  & ...  &  ... &   ...	&  ...	 &  ...	& ...		& ...		& ... 		& 0.44	& ...  \\
?~$\lambda2.174$               &   ...	&  ... 	&   ...	 &  ...	& 0.41	&  ...	   &   ...	 &  ... &   ...	&  ...	 &  0.25	& ...		& ...		& ... 		& ... 		& ...  \\
?{[\ion{Fe}{ii}]}/TiO~$\lambda2.179$      	&  0.49	&  0.22	&   0.47	 &  SL	&   ...	&  ...	   &   ...	 &  ... &   ...	&  ...	 &    ...	& ...		& 0.13A	& ... 		& ... 		&  ... \\
\ion{He}{i}~$\lambda2.182$               &   ...	&  ... 	&   0.38	 &  ...	&   ...	&  ...	   &   ...	 &  ... &   ...	&  ...	 &    ...	& ...		& ...		& ... 		& ... 		&  ... \\
?~$\lambda2.187$               	&  0.19	&  ... 	&   0.50	 &  ...	&   ...	&  ...	   &   ...	 &  ... &   ...	&  ...	 &    ...	& ...		& ...		& ... 		& ... 		& ...  \\
\ion{He}{ii}~$\lambda2.189$     	&  ...		&  ... 	&   0.89	 &  ...	&   ...	&  ...	   &   ...	 &  ... &   ...	&  ...	 &    ...	& ...		& ...		& ... 		& ... 		&  ... \\
?~$\lambda2.193$               &  0.34	&  ... 	&   ...	 &  SL	&   ...	&  ...	  &   ...	 &  ... &   ...	&  ...	 &    ...	& ...		& 0.18A	& ... 		& ... 		&  ... \\
?~$\lambda2.205$               &  0.19	&  ... 	&   ...	 &  ...	&   ...	&  ...	   &   ...	 &  ... &   ...	&  ...	 &    ...	& ...		& ...		& ... 		& ... 		&  ... \\
\ion{Na}{i}~$\lambda2.206$     &  0.61 	&  0.65	&   1.26 	 &  0.33	&   0.68	&  ...	   &  ...	  & 0.73 & 0.39	&  ...	 &   0.59	& 0.54	& 0.11	& 0.83	& 1.08	& 0.41  \\
?~$\lambda2.207$     	&  ... 	&  ...		&   1.65 	 &  ...	&   ...	&  ...	   &  ...	  & ... &  ... 	&  ...	 &    ...	& 0.56	& ...		& ... 		& ...		&  ... \\
\ion{Na}{i}~$\lambda2.209$     	&  0.56 	&  0.68	&   0.83	 &  0.40	&   0.54	&  ...	   &  ...	  & 0.68 &  0.45&  ...	 &   0.53	& ...		& ...		& 0.81	& 1.00	&  0.67 \\
{[\ion{Fe}{ii}]}~$\lambda2.210$			&  0.32	&  ... 	&   ...	 &  SL	&   ...	&  ...	   &  ...	  & ...  & ...	&  ...	 &    ...	& ...		& 0.08A	& ... 		& ... 		&  ... \\
{[\ion{Fe}{iii}]}~$\lambda2.219$	&  0.23	&  ... 	&   ...	 &  ...	&   ...	&  ...	   &  ...	  &  ... &   ...	&  ...	 &    ...	& ...		& ...		& ... 		& ... 		&  ... \\
?H$_{2}$~$\lambda2.225$        		&   ...	&  ... 	&   0.71	 &  ...	&   0.57	&  ...	   &  ...	  &   ...&   0.18	&  ...	 &    ...	& ...		& ...		& ... 		& 0.26 	&  ... \\
?\ion{He}{i}~$\lambda2.228$      	&   ...	&  ... 	&   ...	 &  ...	&   ...	&  ...	   &   ...	 &  ... &   ...	&  ...	 &    ...	& ...		& 0.11A	& ... 		& ...		&  ... \\
?{[\ion{Fe}{ii}]}~$\lambda2.241$ 	&    ...	&  ... 	&   0.59	 &  ...	&   1.50	&  ...	   &  ...	  &  ... &   0.46	&  ...	 &    0.29	& ...		& 0.14	& ... 		& 0.38	&  ... \\
?H$_{2}$/{[\ion{Fe}{ii}]}~$\lambda2.244$ &    ...	&  ... 	&   0.98	 &  ...	&   0.31	&  ...	   &   ...	 &  ... &   ...	&  ...	 &    ...	& ...		& ...		& ... 		& ... 		&  ... \\
{[\ion{Fe}{ii}]}~$\lambda2.254$ 	&    ...	&  ... 	&   0.38	 &  ...	&   0.39	&  ...	   &   ...	 &  ... &   ...	&  ...	 &    ...	& ...		& ...		& ... 		& ... 		&  ... \\
?\ion{He}{ii}~$\lambda2.258$   	&    ...	&  ... 	&   ...	 &  ...	&   0.35	&  ...	   &   ...	 &  ... &   ...	&  ...	 &    ...	& ...		& ...		& ... 		& ... 		&  ... \\
?~$\lambda2.262$              &  ...		& ...		&   0.66 	 &  ...	&   ...	&  ...	   &   ...	 &  ... &   ...	&  ...	 &    ...	& ...		& ...		& ... 		& ... 		&  ... \\
?~$\lambda2.265$      	        &  ...		& ...		&   0.45	 &  ...	&   ...	&  ...	   &   ...	 &  ... &   ...	&  ...	 &    ...	& ...		& ...		& ... 		& ... 		&  ... \\
?~$\lambda2.272$              &  ...		& ...		&   ...	 &  ...	&   ...	&  ...	   &   ...	 &  ... &   ...	&  ...	 &    ...	& ...		& ...		& ... 		& 0.16	&  ... \\
?~$\lambda2.281$              &  ...		& ...		&   0.69	 &  ...	&   ...	&  ...	   &   ...	 &  ... &   ...	&  ...	 &    ...	& ...		& ...		& ... 		& ...		& ...  \\
?~$\lambda2.290$      			        &  ...		& ...		&   ...	 &  ...	&   ...	&  ...	   &   ...	 &  ... &   ...	&  ...	 &    ...	& ...		& ...		& ... 		& 0.22	& ...  \\

\end{longtable}
\end{landscape}
}

\onltab{
\begin{table}
\caption{\label{tbl-6} Line identification and equivalent width measurements for YHGs and Peculiar Oe. Measurements are in units of \AA~and for emission lines unless indicated by an A, indicating the line is in absorption.  Bl indicates a blended line.}             
\centering          
\begin{tabular}{lccc}
\hline\hline
Line & 269723 & 269953 & S124 \\
\hline
?~$\lambda1.955$                                        &  ...          &  ...          &  2.62 \\
?~$\lambda1.976$                                        &  1.10A        &  ...          &  1.36 \\
\ion{Ca}{i}~$\lambda1.978$                       &  1.81A        &  ...          & 0.83  \\
\ion{Ca}{i}~$\lambda1.987$                       &  1.27A        &  .... &  2.27 \\
?~$\lambda1.993$                                        &  0.85A        &  ...          &  1.04 \\
\ion{He}{i}~$\lambda2.058$                       &   ... &  ...          &  45.33\\
\ion{Fe}{ii}~$\lambda2.091$                      &   1.65A       &  ...          & ...   \\
?~$\lambda2.103$                                &   0.21A       &  ...  & ...  \\
?~$\lambda2.107$                                &   0.23A       &  ...  & ...  \\
\ion{He}{i}~$\lambda2.112/3$             &   ... &  ...          &  3.78 \\
?~$\lambda2.120$                                &   0.26A       &  ...  & ...  \\
?~$\lambda2.127$                                &   0.30A       &  ...  & ...  \\
?~$\lambda2.130$                                &   0.13A       &  ...  & ...  \\
\ion{Mg}{ii}~$\lambda2.137$               &   0.53A       &  ...          &  1.28 \\
?~$\lambda2.140$                                &   0.29A       &  ...  & ...  \\
\ion{Mg}{ii}~$\lambda2.144$               &   0.23A       &  ...          & 0.70  \\
\ion{He}{i}~$\lambda2.161$                       & ...  & ... &  3.01   \\
\ion{H}{i}~Br~$\gamma$~$\lambda2.166$ &  3.69A   & 3.40A &  21.46  \\
\ion{He}{i}~$\lambda2.166$                       &  Bl &  Bl    &  Bl   \\
?{[\ion{Fe}{ii}]}/TiO~$\lambda2.179$             &  0.18A        &  ...          &   0.47         \\
\ion{He}{i}~$\lambda2.182$                       &   0.22A       &  ...  &  ... \\
??\ion{He}{ii}~$\lambda2.188$                    &  0.35A        &  ...  & ...  \\
\ion{Na}{i}~$\lambda2.206$                      &  0.22A        &  ...          & 2.54  \\
\ion{Na}{i}~$\lambda2.209$                      &  ...  &  ...          &  2.47 \\
?~$\lambda2.251$                                &  2.22A        & ...           &  ... \\
?~$\lambda2.253$                                &  Bl           & ...           &  ... \\
{[\ion{Fe}{ii}]}~$\lambda2.254$                  &  Bl           &  ...  &  ... \\
?~$\lambda2.256$                                &  Bl           & ...           &  ... \\
?~$\lambda2.257$                                &  Bl           & ...           &  ... \\
?\ion{He}{ii}~$\lambda2.258$                     &  Bl           &  ...  &  ... \\
?~$\lambda2.262$                                &  0.37A        & ...           &  ... \\
?~$\lambda2.265$                                &  0.67A        & ...           &  ... \\
?~$\lambda2.272$                                &  0.77A        & ...           & ... \\
\hline
\end{tabular}
\end{table}
}

\onltab{
\begin{table*}
\caption{Line identification and equivalent width measurements for LBVs. Measurements are in units of \AA~and for emission lines unless indicated by an A, indicating the line is in absorption.  Bl indicates a blended line.}             
\label{tbl-7}
\centering
\begin{tabular}{lcccccc}
\hline\hline
Line &  S52  & S96 & S116 & S128  & S155 & WRAY751 \\
\hline
\ion{Fe}{ii}~$\lambda1.958$          			&  ...		&  5.00	&   ...	 &  4.24	&   ...	&  ... \\
\ion{Fe}{ii}~$\lambda1.976$					&  ...		&  14.84	&   ...	 &  9.46	&   ...	&  ... \\
\ion{Ca}{i}~$\lambda1.978$ 			&  ...		&  ...		&   ...   	 &  0.53	&   ...	& ...  \\
?~$\lambda1.985$					&  ...		&  ...		&   ...	 &  0.44	&   ...	&  ... \\
\ion{Ca}{i}~$\lambda1.987$      			&  ... 	&  4.15	&   ...	 &  3.05	&   ...	&  ... \\
?H$_{2}~\lambda2.002$          			&  ...		&  4.46	&   ...	 &  2.42	&   ...	&  ... \\
{[\ion{Fe}{ii}]}~$\lambda2.008$ 			&   ... 	&  1.15	&   ...	 &  0.59	&   ...	& ...  \\
{[\ion{Fe}{ii}]}~$\lambda2.016$ 			&   ...	& 1.09 	&   ...	 &  0.91	&   ...	& ...  \\
?~$\lambda2.020$               			&   ...	&  0.69 	&   ...	 &  ...	&   ...	&  ... \\
?~$\lambda2.030$               			&   ...	&  0.43 	&   ...	 &  ...	&   ...	&  ... \\
H$_{2}~\lambda2.034$           			&   ...	&  0.17 	&   ...	 &  ...	&   ...	& ...  \\
?\ion{He}{ii}~$\lambda2.037$     		&   ...	&  0.48 	&   ...	 &  ...	&   ...	&  ... \\
{[\ion{Fe}{ii}]}~$\lambda2.046$	 		&   ...	&  1.10 	&   ...	 &  1.33	&   ...	& ...  \\
{[\ion{Fe}{ii}]}~$\lambda2.054$ 			&   ...	&  1.11	&   ...	 &  0.72	&   ...	&  ...	     \\
\ion{He}{i}~$\lambda2.058$     			&   ...	&  6.22	&  0.52	 &  3.25 	&  ...		& 1.54A    \\
\ion{Fe}{ii}~$\lambda2.061$      			&   ...	&  4.80	&   ...	 &  2.89	&   ...	&  ...	  \\
?~$\lambda2.068$               			&   ...	&  ... 	&   ...	 &  ...	&   ...	&  0.24A \\
\ion{Fe}{ii}~$\lambda2.091$     			&   ...	&  15.81	&   0.34	 &  9.22	&   ...	&  ...   \\
?~$\lambda2.097$               			&   ...	&  0.62 	&   ...	 &  ...	&   ...	&  ...	   \\
\ion{He}{i}~$\lambda2.112/3$    		&   ...	&  0.17	&   ...	 &  2.03A	&   ...	&  ...	  \\
\ion{He}{i}~$\lambda2.114$      			&   ...	&  0.15 	&   ...	 &  ...	&   ...	&  ...	    \\
?\ion{N}{iii}/\ion{C}{iii}~$\lambda2.116$ 	&   ...	&  0.38	&   ... 	 &  ...	&   ...	&  ...  \\
?{[\ion{Fe}{ii}]}/\ion{Al}{i}~$\lambda2.118$ &   ...	&  1.31	&   ... 	 &  1.15	&   ...	&  ...	    \\
?H$_{2}$/\ion{Fe}{ii}~$\lambda2.124$ 	&   ...	&  0.41	&   ...	 &  ...	&   ...	&  ...	    \\
{[\ion{Fe}{ii}]}~$\lambda2.133$ 			&   ...	&  0.75 	&   ...	 &  0.69	&   ...	&  ...	   \\
\ion{Mg}{ii}~$\lambda2.137$      		&   ...	&  3.99	&   0.19	 &  2.14	&   ...	&  0.30	    \\
\ion{Mg}{ii}~$\lambda2.144$      		&   ...	&  2.63	&   0.27 	 &  1.69	&   ...	&  0.30	 \\
?{[\ion{Fe}{ii}]}/\ion{He}{i}~$\lambda2.148$ &  ... 	&  0.76 	&   ...	 &  ...	&   ...	&  ...	  \\
\ion{He}{i}~$\lambda2.150$      			&   ...	&  ... 	&   ...	 &  0.56A	&   ...	&  ...	     \\
\ion{He}{i}~$\lambda2.161$      			& ... 		 & 2.42	 &  ...  	&  0.93A  &  ... 		& ...    \\
\ion{H}{i}~Br~$\gamma$~$\lambda2.166$ &   2.88A	 & 84.51 	&  0.60  & 13.78 & 2.93A & 2.20A  \\
\ion{He}{i}~$\lambda2.166$      			&  ... 	&  Bl  	 &  ...  	& Bl       	&  ... 	& ...   \\
?~$\lambda2.169$               			&   ...	&  2.96 	&   ...	 &  ...	&  ... 	& ...  \\
?~$\lambda2.174$               			&   ...	&  0.72 	&   ...	 &  0.22	& ...		&  ...	    \\
?~$\lambda2.187$               			&  ...		&  0.85 	&   ...	 &  ...	&   ...	&  ...	   \\
\ion{Na}{i}~$\lambda2.206$     			&  0.26 	&  0.53	&   ... 	 &  ...	&   0.81	&  1.67	     \\
\ion{Na}{i}~$\lambda2.209$     			&  0.20 	&  0.33	&   ...	 &  ...	&   0.38	&  1.06	    \\
{[\ion{Fe}{iii}]}~$\lambda2.219$			&  ...		&  0.42 	&   ...	 &  ...	&   ...	&  ...	    \\
?H$_{2}$~$\lambda2.225$        		&   ...	&  0.60 	&   ...	 &  0.62	&   ...	&  ...	    \\
?\ion{He}{i}~$\lambda2.228$      		&   ...	&  0.33 	&   ...	 &  ...	&   ...	&  ...	    \\
?{[\ion{Fe}{ii}]}~$\lambda2.238$ 		&    ...	&  0.34	&   ...	 &  ...	&   ...	&  ...	    \\
?{[\ion{Fe}{ii}]}~$\lambda2.241$ 		&    ...	&  2.02 	&   ...	 &  0.50	&   ...	&  ...	    \\
{[\ion{Fe}{ii}]}~$\lambda2.254$ 			&    ...	&  ... 	&   ...	 &  0.58	&   ...	&  ...	   \\
?\ion{He}{ii}~$\lambda2.258$   			&    ...	&  0.79 	&   ...	 &  0.36	&   ...	&  ...	    \\
{[\ion{Fe}{ii}]}~$\lambda2.302$ 			&    ...	&  0.33 	&   ... 	 &  ...	&   ...	&  ...	    \\
?\ion{He}{i}/{[\ion{Fe}{ii}]}/{[\ion{Ni}{ii}]}~$\lambda2.309$ &   1.02	&  ...	 &  ...   &  0.56	&   ...	&  ...	    \\
\ion{Fe}{ii}/\ion{C}{iii}~$\lambda2.325$ 	&     ...	&  0.37 	&   ... 	 &  ...	&   ...	&  ...	  \\
\hline
\end{tabular}
\end{table*}
}

The objects in our survey show a variety of spectral features, with some spectra containing a large number of
emission lines, while others only contain a few absorption lines.  Each spectrum has been analyzed, and 
Tables~\ref{tbl-5}-\ref{tbl-7} contain line identifications and equivalent width measurements for each target star.
Each of the objects in our survey shows evidence of the Br$\gamma$ line
in either emission or absorption.  Because of a close blend with several \ion{He}{i} lines, it is difficult 
to determine the proper shape of the Br$\gamma$. In fact, with the exception of LHA 120-S 124, the Br$\gamma$ line is 
the strongest feature in the spectrum.   All B[e]SGs except LHA 120-S 93 show Pfund lines in emission, as well as the Peculiar Oe star LHA 120-S 124.   
The LBVs S~Dor and LHA 120-S 128 both show Pfund emission, while LHA 115-S 52,
LHA 120-S 116, LHA 120-S 155, and WRAY-15 751 show these lines in absorption.
Neither of the two YHG stars show evidence of Pfund line features.

Examining the spectral features of our collective survey, the groups discussed in the previous section, 
separated according to continua shape, also correspond to distinct spectral compositions.  Stars in the first group, 
characterized by a kinked continuum, contain few, primarily absorption lines in their spectra, including lines 
of \ion{H}{i}, \ion{Na}{i}, and \ion{Mg}{ii}, typical of cooler (A-, F-, or G-type) supergiants.  Spectra in the 
second group, characterized by a generally flat continuum, contain a large number of emission lines 
 (e.g., \ion{H}{i}, \ion{He}{i}, \ion{Fe}{ii}, [\ion{Fe}{ii}], \ion{Mg}{ii}, \ion{Na}{i}, and/or 
\ion{Ca}{i}), with some targets showing 
lines in absorption as well.  These spectra are more representative of hotter (O- or B-type) supergiants.
LHA 120-S 12 shows in several of its spectral features what appear to be shell line 
profiles, possibly due to its edge-on orientation \citep{Aret}.  In several objects, helium lines 
are found in absorption, consistent with a B-type stellar spectrum. 

Out of the 25 sample stars, we detected CO band head emission in a total of 13 objects, of which 10 are classified
as B[e]SGs, 2 as YHGs, and 1 as a Peculiar Oe star (see Fig.~\ref{HRplot2}.  For three of these stars, 
LHA 115-S 65 \citep{Oksala12}, MWC 137 \citep{Muratore13} and 
LHA 120-S 35 (Torres et al., in preparation), this study is the initial detection of CO emission.  
The remaining 10 stars are confirmed detections, previously known to show CO in emission.  
The spectrum of LHA 120-S 59 shows evidence for weak CO emission, although the contamination of 
strong Pfund emission makes this determination unclear.  If present, it would be the first detection
of CO in this star.  We find that none of the LBVs show evidence of 
CO molecular emission.  Although our sample is small, it seems that the lower metallicity of the SMC 
compared to Galactic metallicity does not seem to have an effect on the presence of CO emission.  
However, all three of the observed SMC B[e]SGs from this sample are highly luminous, and we cannot rule out 
an observational bias.  It is interesting though to note that there does appear to be a lower luminosity limit, with 
no stars lower than $\log$ L/L$_{\odot}$ = 5.0 showing definitive molecular band head emission.  Of the LMC B[e]SGs, 
we find 6 of the 10 do not show CO emission, three of which are below this ``luminosity limit'' 
(LHA 120-S 59, LHA 120-S 93, and LHA 120-137).

\section{Spectral modeling}\label{models}

Visual inspection of the region between 2.27 and 2.4 $\mu$m reveals that many of the observed objects
contain evidence of circumstellar material either in the form of hydrogen recombination emission lines
from the Pfund series, or first overtone band head emission from CO molecules. 
To determine the physical parameters of the CO emitting region, the contribution from other sources was removed.
Using the spectral region between the Br$\gamma$ line and the start of the CO band heads, we determined a suitable 
linear fit to the continuum using the IRAF task \textit{continuum} for each target.  This linear fit was then 
subtracted from the spectrum to determine the accurate CO contribution to the flux.  This spectrum was 
compared with synthetic spectra to determine the best fit model, using the codes described in the 
following sections.  We find a variety of modeling situations, with some spectra requiring both CO and Pfund models, 
while some only require one or the other.  When only Pfund emission lines are visibly present, the spectra were modeled 
accordingly, and inspected for any trace of CO emission.  In the case of no spectral features or Pfund lines 
in absorption, we do not compute a model.

\subsection{Pfund model}

The Pfund series emission spectrum was in most cases modeled with the code developed by \citet{Kraus00}, 
with the spectral lines computed using a simple Gaussian profile.  The emission lines are assumed
to be optically thin and in LTE.  This model is applicable in the case that the emitting material 
originates in a wind or a shell.  In a few cases, we noted that the observed Pfund emission lines 
were not well fit by a Gaussian and appeared to be double peaked.   
We therefore developed a new code, similar to the CO emission code of \citet{Kraus00}, to consider emitting 
material located in a narrow ring within a rotating disk. 

\begin{figure*}[!ht]
\centering
\begin{tabular}{cc}
\includegraphics[width=80mm]{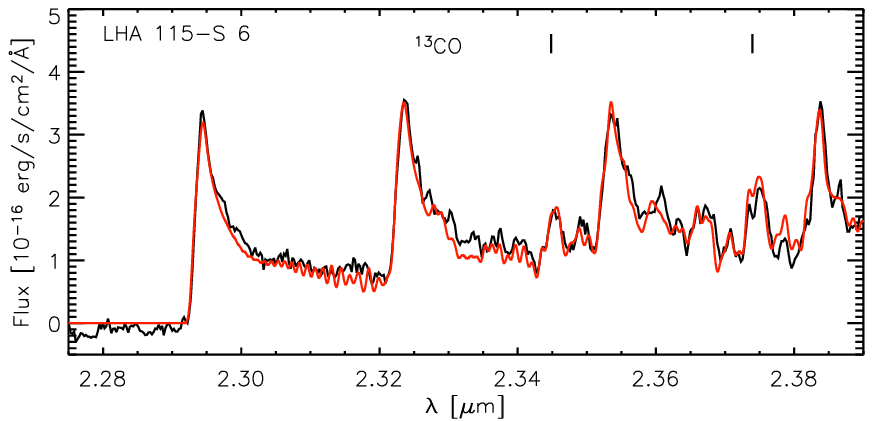} &
\includegraphics[width=80mm]{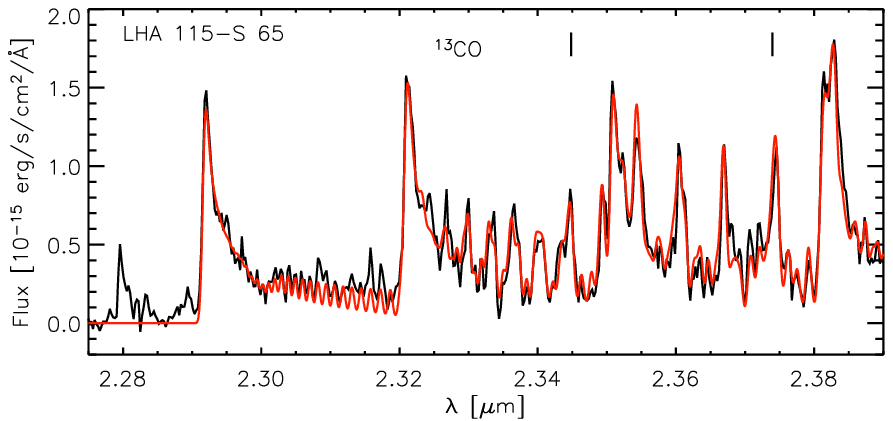} \\
\includegraphics[width=80mm]{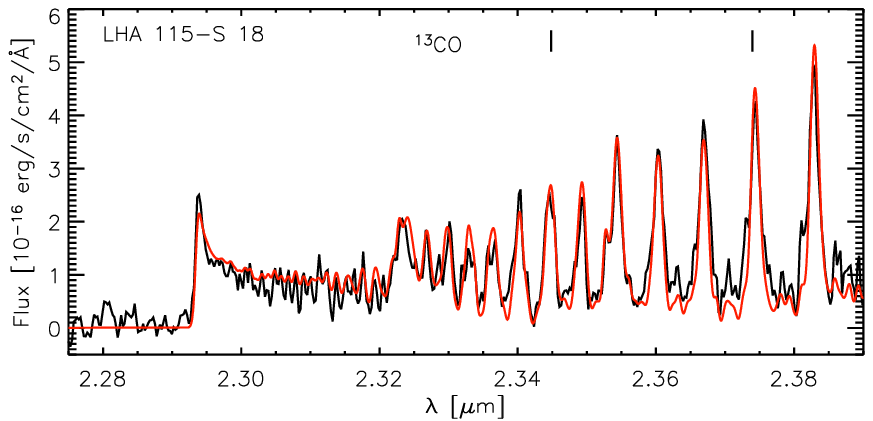} &
\includegraphics[width=80mm]{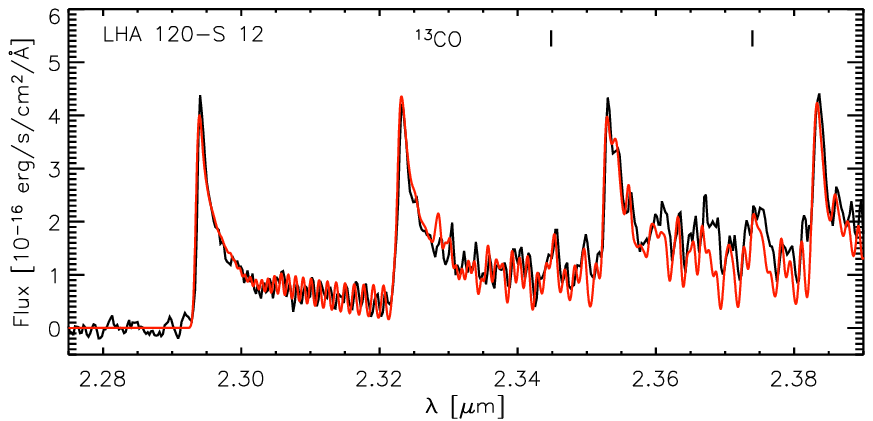} \\
\includegraphics[width=80mm]{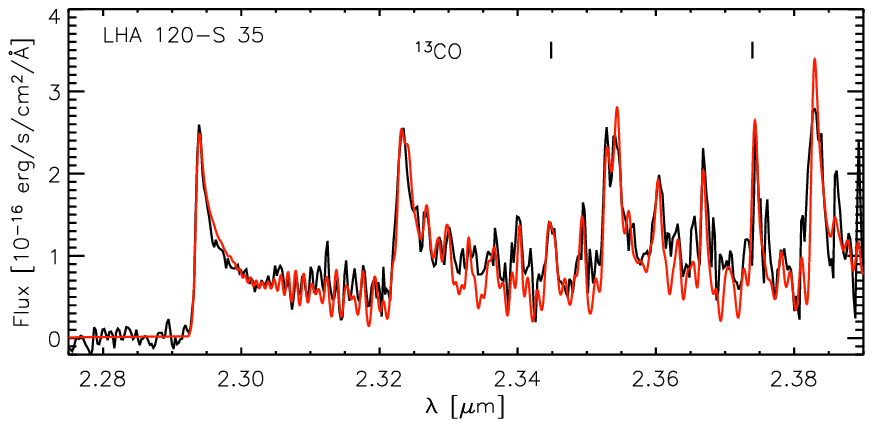} &
\includegraphics[width=80mm]{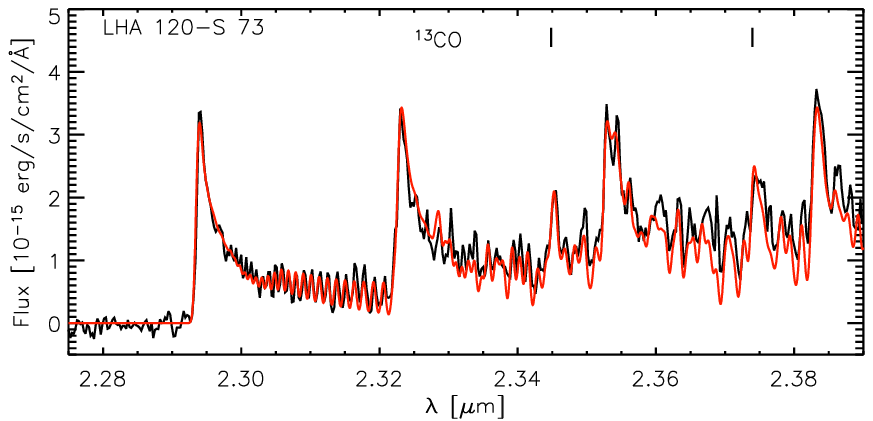} \\
\includegraphics[width=80mm]{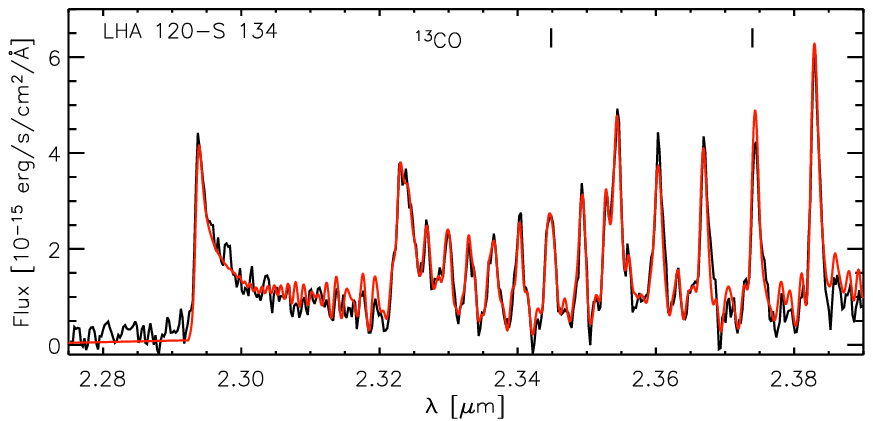} &
\includegraphics[width=80mm]{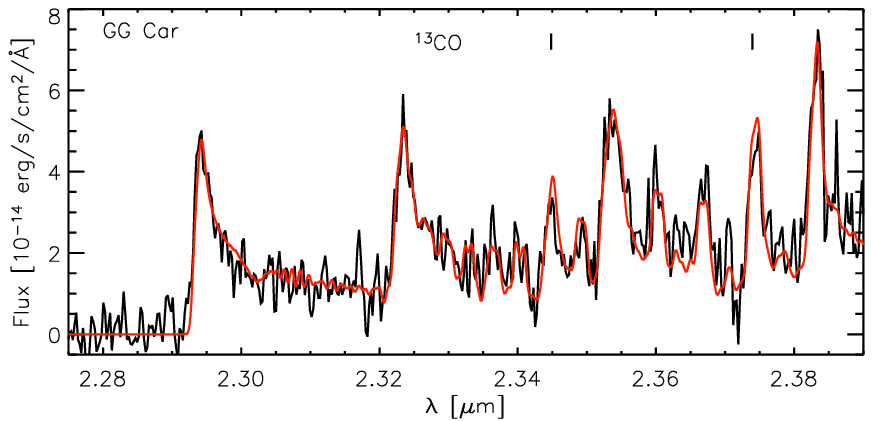} \\
\includegraphics[width=80mm]{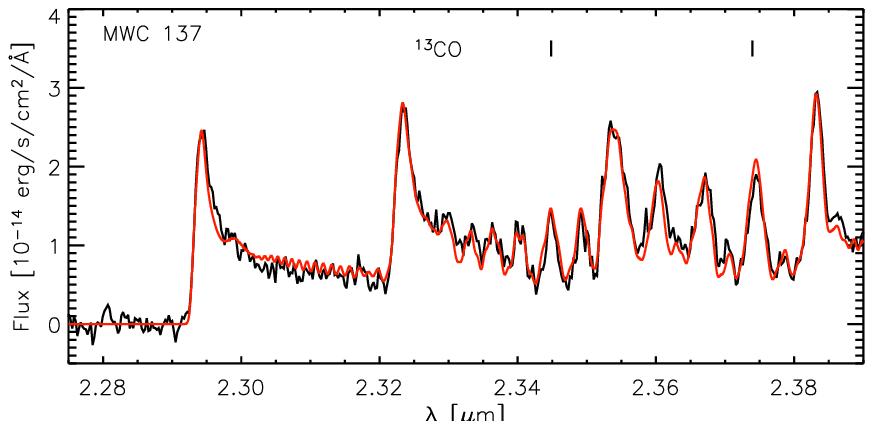} &
\includegraphics[width=80mm]{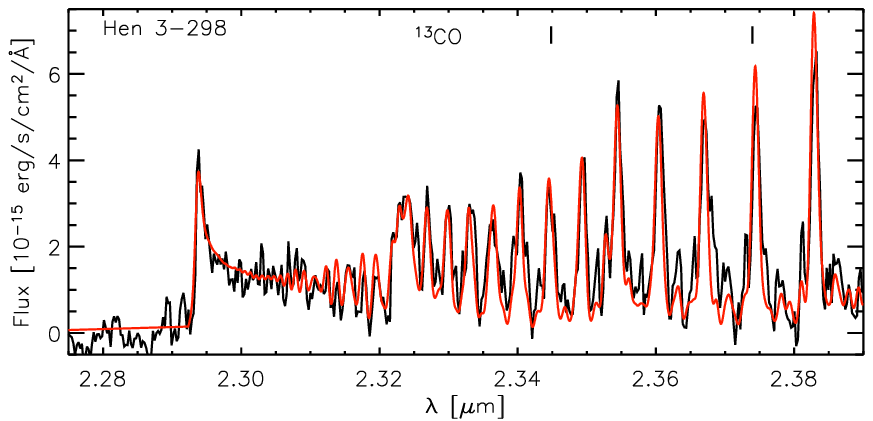}
\end{tabular}
\caption{Model fits (red) to the flux-calibrated, continuum subtracted CO band spectra (black) of B[e]SGs.  
The positions of the $^{13}$CO band heads are indicated. }
\label{COpresSG}
\end{figure*}

\subsection{CO model}

For the model computations of CO band head emission, we utilized the previously developed 
code of \citet{Kraus00}, which computes a $^{12}$CO emission spectrum with 
the material located within a small ring in a rotating disk with lines broadened 
by a rotational velocity $\varv_{\rm{rot}}$, and the code of \citet{Kraus09} which computes a non-rotating 
spectrum of both $^{12}$CO and $^{13}$CO band head emission.  In both cases the CO gas is assumed to be 
in LTE.  Since several of the objects with CO emission appeared rotationally broadened 
(beyond the SINFONI velocity resolution of 60~km~s$^{-1}$), 
but with significant $^{13}$CO emission, we used the codes of \citet{Kraus00} and \citet{Kraus09} as the
base for a hybrid model that computes both the $^{12}$CO and $^{13}$CO emission spectra assuming that 
the emitting material is located in a thin, rotating ring within the circumstellar environment.  
The assumption that the material is located in a relatively narrow ring around the star with constant 
temperature, column density, and rotational velocity is supported by 
both theory \citep[see e.g.,][]{Kraus09} and observations \citep[see e.g.,][]{Kraus00,Liermann,Kraus13}.  
It may be possible that the material is distributed in a much different configuration, such as a shell 
or bipolar LBV type ejecta.

\subsection{Model results}\label{modelresults}

\begin{table*}[!ht]
\centering
\caption{Best-fit CO model parameters} 
\label{tbl-3}
\begin{tabular}{lcccc}
\hline\hline
Object & T$_{\rm{CO}}$ & N$_{\rm{CO}}$ & $^{12}$C/$^{13}$C & A$_{\rm{CO}}\cos i$ \\
 & (K) & (10$^{21}$ cm$^{-2}$) &  & (AU$^{2}$) \\
\hline
LHA 115-S 6    & 2200$\pm$200 & 5$\pm$2      & 12$\pm$2  &  1.00$\pm$0.08  \\
LHA 115-S 18   & 2000$\pm$200 & 8$\pm$3      & 20$\pm$5  &  8.9$\pm$0.6  \\  
LHA 115-S 65   & 3200$\pm$300 & 1.5$\pm$0.5  & 20$\pm$5  &  11.7$\pm$0.9 \\  
HD 269953      & 3000$\pm$200 & 2$\pm$0.5    & 10$\pm$1  &  41.5$\pm$1.1 \\  
LHA 120-S 12   & 2800$\pm$500 & 2.5$\pm$0.5  & 20$\pm$2  &  2.33$\pm$0.06  \\  
LHA 120-S 35   & 3000$\pm$200 & 2$\pm$0.5    & 10$\pm$2  &  1.16$\pm$0.06  \\ 
LHA 120-S 73   & 2800$\pm$500 & 3.5$\pm$0.5  & 9$\pm$1   &  17.8$\pm$0.5 \\  
LHA 120-S 124  & 3000$\pm$400 & 9.5$\pm$0.5  & 20$\pm$5  &  0.90$\pm$0.02 \\  
LHA 120-S 134  & 2200$\pm$200 & 2$\pm$1      & 15$\pm$2  &  62$\pm$14  \\
MWC 137        & 1900$\pm$200 & 3$\pm$2      & 25$\pm$5  &  2.04$\pm$0.27  \\ 
GG Car         & 3200$\pm$200 & 5$\pm$3      & 15$\pm$5  &  0.58$\pm$0.05 \\ 
Hen 3-298      & 2000$\pm$200 & 0.8$\pm$0.4  & 20$\pm$5  &  11$\pm$4 \\
\hline
\end{tabular}
\end{table*}

\begin{figure}[!ht]
\centering
\begin{tabular}{c}
\includegraphics[width=80mm]{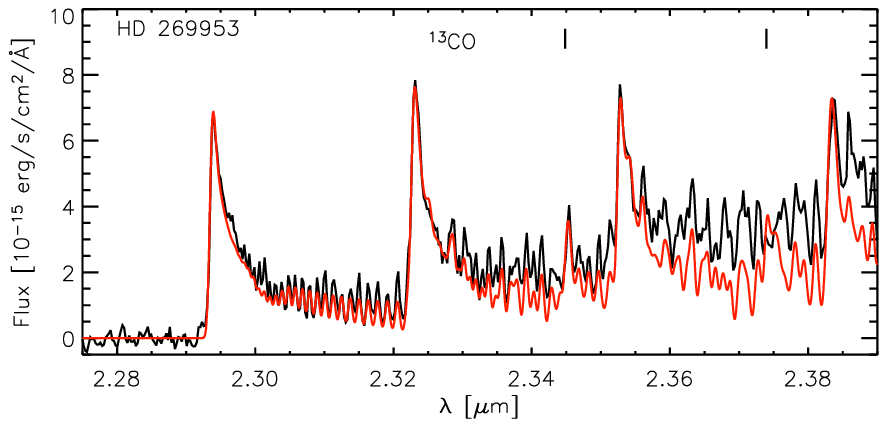} \\
\includegraphics[width=80mm]{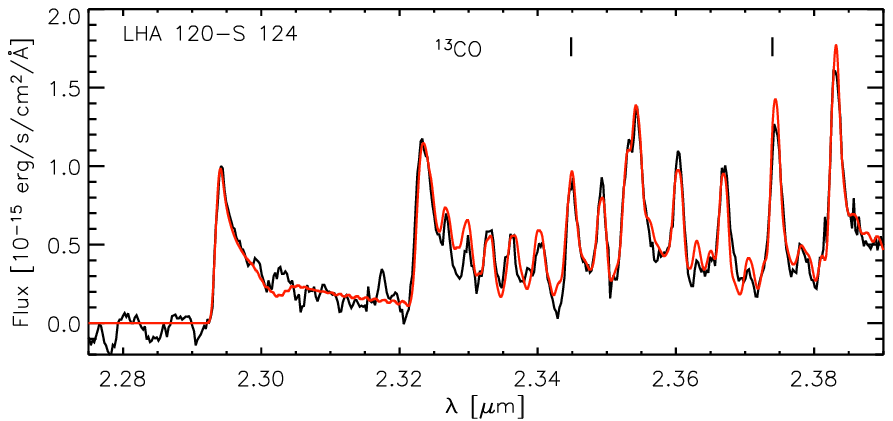} 
\end{tabular}
\caption{Model fits (red) to the flux-calibrated, continuum subtracted CO band spectra (black) of the 
YHG HD 269953 and the Peculiar Oe star LHA 120-S 124.  The positions of the $^{13}$CO band heads are indicated.}
\label{COnonSG}
\end{figure}

\begin{figure}[!ht]
\centering
\begin{tabular}{c}
\includegraphics[width=80mm]{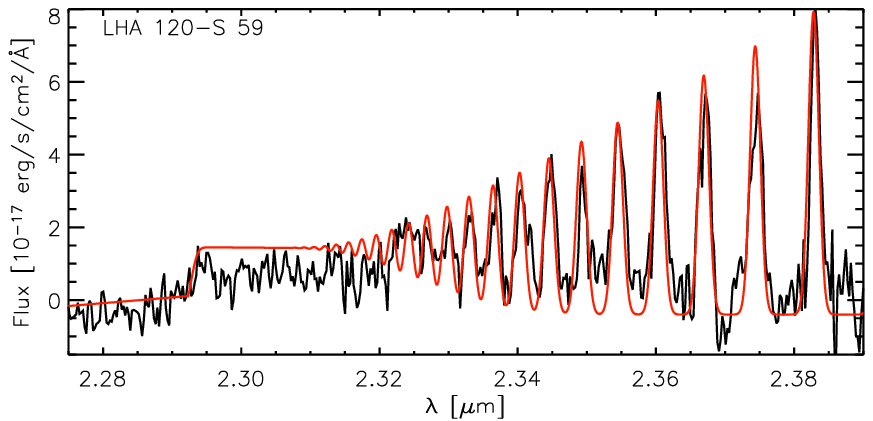} \\
\includegraphics[width=80mm]{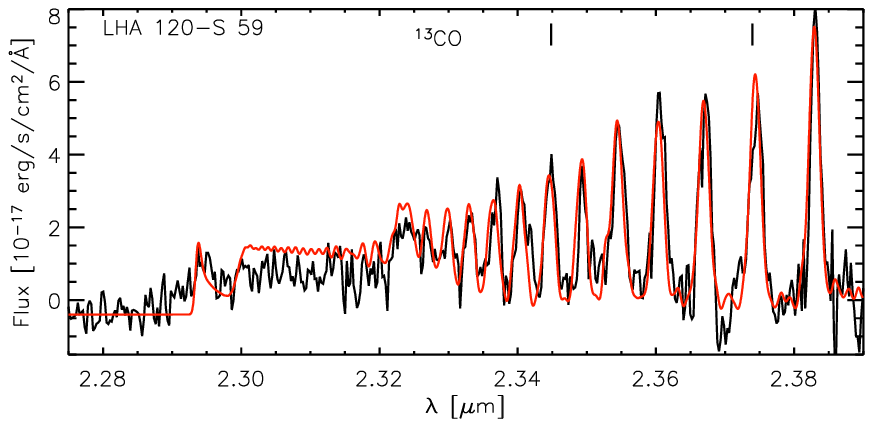}
\end{tabular}
\caption{Model fits (red) to the flux-calibrated, continuum subtracted CO band spectra (black) of the 
B[e]SG LHA 120-S 59.  It is unclear whether a pure Pfund emission model (\textit{top}) or a weak CO 
plus Pfund model (\textit{bottom}) is the physically appropriate model. The positions of the $^{13}$CO 
band heads are indicated. }
\label{COpresLHA59}
\end{figure}

Model computations encompassed a large range of temperatures and column densities.  
The models have each been computed using the codes described 
in the previous sections, with only three stars (MWC 137, GG Car, and LHA 115-S 6) requiring an additional rotational broadening greater than 
the instrumental broadening of 60 km s$^{-1}$.  
The best fit models to each of the targets with CO band head emission and/or Pfund emission are plotted in red in 
Figs.~\ref{COpresSG}-\ref{COpresLHA59}, except for HD 269723 (see the explanation below).  
The spectra shown in these figures have been continuum subtracted to 
determine the accurate CO contribution to the flux (as described in the beginning of this section).  
In the case of LHA 120-S 59 (Fig.~\ref{COpresLHA59}), two 
models are presented, one with weak CO emission and one without.  Both models fit the observations relatively well, 
however, the quality of the spectrum makes differentiation impossible.  

Tables~\ref{tbl-3} and \ref{tbl-4} 
give the model parameters for each of the CO emission fits.  We do not include the model parameters for HD 269723 
in this table, as the parameters could not be determined accurately.  One possible reason for this difficulty may 
be that, as a YHG, the star might be surrounded by a cool molecular shell, rather than a disk. 
Also excluded is the model of LHA 120-S 59 
as it is merely suggestive at this point.  The last column of Table~\ref{tbl-3} gives a line of sight
computed area for the emitting ring of CO, A$_{\rm{CO}}\cos i$, in units of AU$^{2}$. This quantity is derived from
the ratio of the model fit to the flux calibrated spectrum.  The area calculation assumes that each 
target's distance has been determined.  For the LMC and SMC, these values are relatively well know at 48.5 
and 61 kpc, respectively.  The galactic objects have much less certainty in their distance 
determinations.  For GG Car, \citet{Marchiano} determine a distance of 5 $\pm$ 2 kpc.  \citet{EF98} report a 
lower limit of 6 kpc for the distance to MWC 137, assuming that the star is a supergiant.  
A range of 3-4.5 kpc for Hen 3-298 was reported by \citet{Miro05}; we use the upper limit for our calculation, 
which results in an upper limit for the emitting area.

The objects in our survey with detected CO represent a range of L and T$_{\rm{eff}}$ values, 
however, there do not seem to be any correlations between stellar properties and the conditions in the 
circumstellar material where the CO molecules are formed.  As CO band emission typically arises from 
the inner rim/edge of the molecular disk region, the fact that our CO model temperatures
are all much cooler than the dissociation temperature of 5000 K indicates that the 
material may be located in a detached disk structure, as opposed to a disk that reaches the stellar surface as 
suggested by the model of \citet{Zickgraf86}.  Such behavior has recently been reported in other studies
\citep{Kraus10,Liermann,Wheel,Cidale12,Kraus13}.

\begin{table}[!ht]
\centering
\caption{Broadening parameters used in model computation}
\label{tbl-4}
\begin{tabular}{lccc}
\hline \hline
Object & $\varv_{\rm{rot,CO}}~\sin~i$ & $\varv_{\rm{rot,Pf}}~\sin~i$  & $\varv_{\rm{gauss,Pf}}~\sin~i$ \\
  & (km~s$^{-1}$) & (km~s$^{-1}$) & (km~s$^{-1}$)  \\
\hline
LHA 115-S 6     & 100$\pm$10  &  130$\pm$10  & --- \\
MWC 137         & 85$\pm$10  & ---  &  140$\pm$40  \\ 
GG Car\tablefootmark{a}          & 80$\pm$1  &  94$\pm$2 & --- \\ 
\hline
\end{tabular}
\tablefoot{Velocities determined by other work:
\tablefoottext{a}{\citet{Kraus13}}}
\end{table}

Notably, one of the most engaging results from this study is 
that each of the targets with detectable $^{12}$CO band head emission also shows clear signs of $^{13}$CO emission.  
From the investigation by \citet{Kraus09}, it is clear that pre-main sequence levels of $^{13}$C should not 
produce visible molecular band head 
emission.  We therefore conclude that each of the objects in this study with detected CO emission is in fact 
an evolved, post-main sequence object.  This point is especially poignant in the case of MWC 137 which has been 
frequently cited as a possible Herbig B[e] star \citep{Hill92}.

\section{Discussion}

\subsection{Evolutionary status determination}

\subsubsection{LBVs}\label{LBVstate}

Although our survey contains only a small number of 
LBV samples, it appears that there is clearly a dual 
dichotomy as far as spectral morphology is concerned. 
In fact, the overall continuum shape combined with
the spectral line constitution is a clear sign
of the physical state and the evolutionary status of such stars. 
If the spectrum appears with many emission lines and a relatively
flat continuum, the star will be in a more blue
ward position in the HRD, similar in appearance
to the B[e]SGs. These LBVs are in their
quiescent state and are not currently experiencing
heavy or violent mass-loss events. Only two LBVs in
our sample fall into this category, S~Dor and LHA 120-S 128. 
In addition, these two stars happen
to be the most luminous of our LBV sample, and
are located near the Humphreys-Davidson limit
in the HRD (see Fig.~\ref{HRplot2}).   Based on the classification scheme
presented by \citet{Clark} for LBVs, the strong presence of \ion{He}{i} 2.058 $\mu$m
indicates that both of these stars are quite hot. 

The remaining LBVs in our sample that display
the double kinked, YHG-like continuum with few, mostly absorption lines
all have lower luminosities than their B[e]SG-like 
counterparts, and happen to appear in 
their cooler HRD position.  The spectral line constitution of this group of stars is very different from 
the previous group, with some diversity among the members.  Their appearance indicates
cooler spectral types (late B and later).  The strong \ion{Na}{i} doublet emission in WRAY 15-751 and LHA 120-S 116, 
combined with \ion{He}{i} absorption and weak \ion{Mg}{ii} emission suggests a later B spectral type.  These stars are currently 
in the so-called `active' state of their evolution.
Outbursts have been reported recently for both WRAY 15-751
\citep{Sterken} and LHA 120-S 155 \citep{Gamen09}, confirming their active
status.  
Further, the remarkable brightening we observe with respect to the original 
2MASS magnitude measurements (see Sect.~\ref{Kmag}) reinforces this conclusion.
Both LHA 115-S 52 (S 52) and LHA 120-S 116 lack sufficiently recent observations to confirm
or deny this status, however the observed brightening of S~52 is suggestive. 
The study of \citet{Massey07} 
for emission objects in M31 and M33 further corroborates these two spectral
categories, labeling them ``hot'' and ``cool'' LBVs, based on their
optical appearance.  

Whether these low luminosity LBVs are 
currently in a post-RSG phase is hard to tell,
as they all appear to have similar initial masses 
of $\sim 40$\,M$_{\odot}$, the boundary mass noted by
\citet{Meynet11}, below which stars develop into LBVs
as post-RSGs.  More massive stars become LBVs directly
after the main sequence.

\begin{figure*}[!ht]
\centering
\includegraphics[width=90mm, angle=-90]{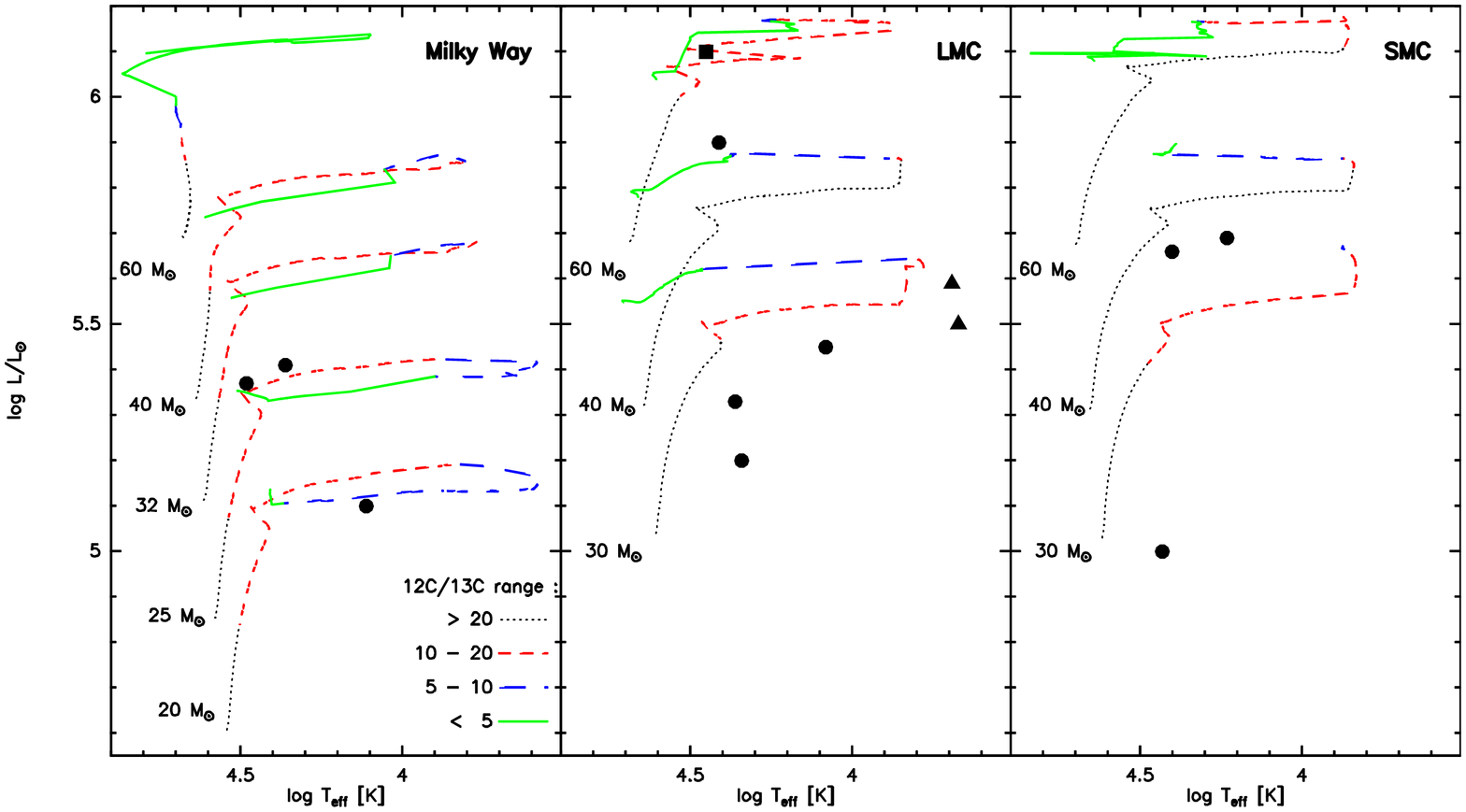}
\caption{HR Diagram showing the location of each target star with detected CO emission.  The evolutionary tracks 
from \citet{MM05} (LMC, SMC) and \citet{Ekstrom} (Milky Way) trace the change in $^{12}$C/$^{13}$C ratio with stellar evolution.  For the 
Galaxy, the tracks are computed for an $\Omega = 0.4$, while the Magellanic Cloud tracks are for 
$\varv_{\rm{eq}} = 300$ km~s$^{-1}$.  $\Omega$ is the ratio of the rotational velocity to the critical rotation 
rate and $\varv_{\rm{eq}}$ is the equatorial rotational velocity.  The symbols are explained in the 
caption of Fig.~\ref{HRplot2}.}
\label{evol_rot}
\end{figure*}

\subsubsection{B[e]SGs and other objects}\label{statusBe}

As a group, the spectra of B[e]SGs fall into two categories based on spectral line composition.  A majority of the objects display 
a hot spectrum, with strong helium and metal lines indicative of an early B spectral type.  The remaining objects, LHA 120-S 73,
LHA 120-S 35, LHA 120-S 59, and LHA 120-S 137, are likely later B supergiants, displaying fewer spectral lines, with some lines observed 
in absorption.  These general assessments align with previous spectral type determinations from optical spectra as reported in
Table~\ref{tbl-1}.

One of the main results of the work presented in this paper is the comprehensive determination of the 
$^{12}$C/$^{13}$C ratio for a variety of objects. While 
we cannot judge the evolutionary status of B[e]SGs that do not show any CO features, the 
presence and strength of the $^{13}$CO molecular emission may be a tool to distinguish the ambiguity between objects that 
resemble both Herbig Ae/Be stars and B[e]SGs \citep[shown previously by][]{Liermann, Muratore}.

The model derived ratios listed in Table~\ref{tbl-3},
combined with the evolutionary tracks from \citet{MM05} and \citet{Ekstrom} shown in Fig.~\ref{evol_rot}, 
provide evidence for the evolutionary status of each of these objects.  Note that these tracks are for rotating models,
as described in the figure caption.  Identical tracks for non-rotating models show that the $^{12}$C/$^{13}$C ratio
does not drop below 20 until after the RSG phase, and reaches values $\leq$ 5 for those tracks 
returning to the blue side of the HRD.  Given the range of values 
observed in our objects, the non-rotating scenario does not seem to apply.  The sample targets must therefore
be rotating to achieve the surface chemical enrichment observed.

Although there are not as many tracks 
for the LMC and SMC objects, we can still get a preliminary idea of their evolutionary location based 
on the $^{12}$C/$^{13}$C ratio. The majority of B[e]SGs appear, based on their $^{13}$C enrichment, 
to be pre-RSG phase objects.  However, the three stars with the lowest $^{12}$C/$^{13}$C ratios, LHA 115-S 6, 
LHA 120-S 35, and LHA 120-S 73, may be close to entering the RSG phase or perhaps evolving blueward after their 
RSG phase.  This distinction is difficult to ascertain without more information on the rotational speed 
of the star and/or without more evolutionary information for SMC and LMC metallicity stars.  LHA 115-S 6 is a 
distinct case among these B[e]SGs that has been suggested to be the result of a binary merger.

As expected, the YHG HD 269953 is confirmed to be a post-RSG object.  
Although we are unable to attain an accurate value for HD 269723, we may assume that it too is in a 
post-RSG phase.  The LMC Peculiar Oe star LHA 120-S 124 (S 124), which has also been suggested to 
be an LBV \citep{vanG} and a peculiar O supergiant \citep{SW87},
can be presumed to have left the main-sequence, but not evolved past the RSG phase.   
S 124 could be considered an LBV candidate, due to its 
photometric variability, however, its $K$-band spectrum, and by extension its circumstellar environment, is much different 
than that of ``bonafide'' LBVs.  Based on its stable CO emission, 
previously detected by \citet{McGregor88b}, the star more similarly resembles a B[e]SG, and could 
be considered a slightly hotter counterpart to these objects.

\subsection{Incidence of CO emission in B[e]SGs}

As a group, the B[e] stars present similar spectral features, but with varying physical parameters 
and evolutionary stages \citep[see e.g.][]{Lamers98}.  As an example, the B[e]SGs seem to be quite 
heterogeneous in the mass-loss traced by their circumstellar environments,  
with presumably the same mechanism forming their disk-like structure.  While they may differ in mass and 
temperature, each object shows the same hybrid spectrum, forbidden emission, and infrared excess due to outer
dust regions \citep[e.g.,][]{Zickgraf86, Bonanos09, Bonanos10, Kastner10}. One would thus expect 
that the $K$-band spectra, which contains information on the conditions
of the circumstellar material, would be quite similar for each of these objects, and in the presence 
of dust, the disk structure would also form molecules.  The result of this work that not all B[e]SGs 
show CO emission is therefore surprising.

\begin{figure}
\centering
\includegraphics[width=90mm]{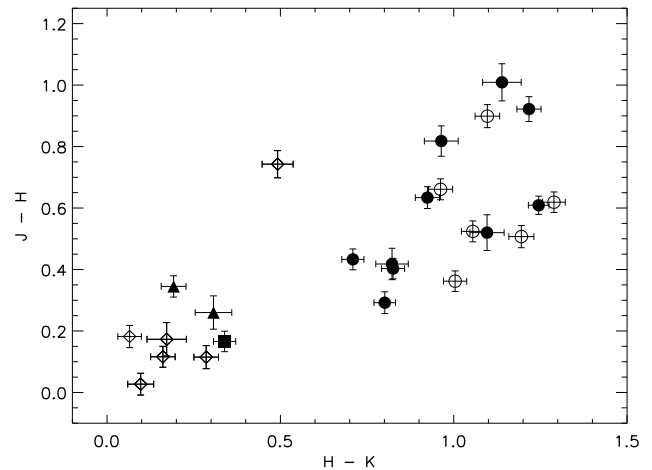}
\caption{Color-color diagram plotting each target's J-H value and H-K value calculated from the corresponding
JHK magnitudes listed in the 2MASS point source catalog \citep{Cutri}.  The symbols are explained in the 
caption of Fig.~\ref{HRplot2}.  The B[e]SGs are all located on the right side of the figure, indicating 
the presence of dust.  The other objects do not show such a large excess, with the exception of WRAY 15-751, 
which stands outside both groups due to a cool companion.}
\label{JHHKplot}
\end{figure}

As evidenced by the HRD displayed in Fig.~\ref{HRplot2}, there seems to be a lower luminosity 
limit as previously discussed.  This may be due to lower mass-loss rates for these stars, leading to 
circumstellar environments which are of too low density to permit the detection of molecular emission.  
But, this limit does not 
explain the absence of CO emission in the higher luminosity stars LHA 120-S 127, LHA 120-S 89, and LHA 120-S 22.
These stars appear in the HRD nearby to stars which show evidence for CO molecules.  In the color-color diagram
shown in Fig.~\ref{JHHKplot}, each non-CO emitting star is coincident with a B[e]SG with CO.  B[e]SGs 
in the Galaxy, in the LMC, and in the SMC all show CO emission, eliminating metallicity as a determining factor.  
Our results may seem to suggest that only a portion of stars in the LMC do not show CO, but this should be seen as a 
small statistics issue, a problem only solved with the discovery of more group members.

We can speculate on the conditions which may be different in each of the stars in the sample without 
CO emission.  One possibility is the structure of the circumstellar material may be different, although B[e]SGs are 
presumed to possess dense disks, excellent locations for molecular formation, the current state of polarization 
studies only indicate that in a majority of these stars the material is not spherically distributed 
\citep{Magalhaes92,OD99,Melgarejo,Magalhaes06}.  \citet{Aret} find, based on Ca and O emission lines 
formed in the inner region of the circumstellar region, a spiral structure of material for LHA 120-S 22.  
The absence of hot CO emission could be related to the conditions of the material in the region where 
these molecules are formed, which may be at too low of a temperature or too low of a density for emission
features to be detected in the spectrum.  To investigate this, it would be prudent to obtain observations 
in different wavelength ranges to study the molecular content of the disks.

Another conceivable explanation is that these stars are not B[e]SGs, but instead may be stars 
experiencing LBV conditions.  For example, LHA 120-S 127 (S 127) is the brightest B[e]SG in our survey, 
and lies in the HRD very near to the Humphreys-Davidson limit \citep{HD94}, similar in position 
to the known LBVs LHA 120-S 128 and S Dor.  As S 127 is viewed pole-on with respect to its 
circumstellar material \citep[e.g.,][]{Aret}, observations should reveal the entirety of the 
material composition.  While studies show that S 127 has strong Ca and O emission in the inner region of its 
circumstellar environment \citep{Aret} and copious amounts of dust further out \citep{Bonanos09, Kastner10}, no trace 
of CO emission is detected in our survey.  As CO molecules form at a location intermediate to the 
Ca and O emitting material and the dust particles, the obvious conclusion is that the material 
surrounding S 127 is not a continuous disk, but a series of rings or shells.  This type of structure 
is contrary to the hybrid wind model for B[e] stars and suggests episodic mass-loss events as in the 
case of S Dor ejections.  

The role of binarity in the formation of disks and rings confined to the equatorial plane of evolved massive stars 
remains unclear.  In the case of GG Car, its binary companion is close enough to interact, and the 
CO molecules are located in a circumbinary ring around the system.  The origin of the material is 
likely from a main sequence remnant Be disk, but there is a small possibility of it being due to Roche 
lobe overflow (RLOF) due to the two stars close orbit \citep{Marchiano,Kraus13}.  
The Galactic B[e]SG HD 327083 \citep{Wheel} is another example of a 
system with circumbinary material similar to GG Car, as well as detected CO emission.  
In these two cases, the stable presence of the CO emission and the appearance of the material in
a detached ring, may be a consequence of the binary interaction.  As this study finds detached circumstellar structure,
evidenced by cooler CO emitting gas, it would be straightforward to make the conclusion that all stars with CO emission
exist in close binary systems.  However, there are several 
counter points to disparage binarity as a factor in the formation of disks or rings.  The star MWC 137 has 
no visible companion \citep{MarMc08}, and yet shows detectable $^{12}$CO and $^{13}$CO.  LHA 115-S 18 was suggested 
to have a binary companion, but no observational evidence backs up such a claim \citep{Torres}.  Similarly, 
CPD-52 9243 \citep{Cidale12} has been suggested to be a binary, but there has been no confirmation.  LHA 115-S 6 
has a binary companion, however, the orbital distance is large and there is no possibility of RLOF interaction 
between the two objects \citep{Zickgraf96}.  No binarity has been indicated for any of the other 
CO-emitting stars, and thus at this point we are hesitant to place any confidence in the role of binarity as a 
circumstellar disk formation and maintenance mechanism.  More work in this aspect of B[e] supergiant research 
may prove incredibly insightful with the increasing capabilities of interferometry.

\subsection{Influence of rotation on stellar mass loss}\label{rotation}

 The detection of Keplerian rotating disk material around some B[e]SGs 
\citep{Kraus10,Wheel,Cidale12} is in direct opposition to the outflowing disk expected 
by the model of \citet{Zickgraf85}.  Further, the general stability of these disks, and in particular the presence of dust 
particles, indicate that the material is not in constant motion, as in an outflowing disk.  While this initial model 
may not be physically applicable, rotation, more specifically rapid-rotation, may play a large role in the details
of the stellar mass loss.  

Viscous disks have been studied in the context of classical Be stars and suggested 
for the case of S 65 \citep{Oksala12}. 
This star is particularly interesting since the CO emission appears to be traveling at a higher outward speed than other parts of the 
circumstellar environment.   The CO band heads required an additional 
shift in velocity of 196 km s$^{-1}$ to simultaneously fit the CO band head emission and the Pfund emission.
This star is a confirmed rapid rotator \citep{Kraus10}, rotating at a minimum of 75\% of 
the critical speed.  Could the rapid stellar rotation be the cause of this high velocity ejection?  S 65 is 
thought to be viewed disk edge on, and thus it could be rotating at a nearly critical velocity, considering 
the $\varv~\sin~i$ measurement was obtained from photospheric line profiles, which are unable to determine rotation 
speeds higher than 75\% of critical \citep{Town}.  If we consider that the star may be critically rotating, 
it could have ejected a large amount of mass via the process described by \citet{Kurfurst}, 
with the star subsequently spinning down due to the large loss of angular momentum.  While this situation has
potential, the only way to determine the physical explanation is new observations.  The possibility also 
still exists that this object is an LBV and we are viewing the result of an outburst. 

 LHA 120-S 73 is another star with a determined rotation velocity ($\varv~\sin~i \sim 50$~km s$^{-1}$).  \citet{Aret} 
found the star may be oriented more or less pole-on with $i = 28^{\circ}$.  This would indicate a rotation at 
$\sim$75\% critical, nearly identical to S 65.  Besides the cause of the mass-loss, another 
possible effect of rapid rotation
in these stars is the composition of the ejected material.  As the rotation rate increases, the 
chemical enrichment achieved during evolution is mixed and transported to the stellar surface at a 
higher rate.  Two similarly evolved stars, one with a much higher rotation rate, may have very different 
$^{12}$C/$^{13}$C ratios.  This could be the explanation of the higher ratios found in S 73 and S 35 (although the 
$\varv~\sin~i$ of this star has not been determined).  An alternative 
explanation presumes that the B[e]SGs are all in the post-RSG phase, with the stars with lower amounts of  $^{13}$CO 
 initially slowly or non-rotating.  This scenario is much less likely considering the blueward location of these objects 
in the HRD, and the known rotational velocity of S 65.

\subsection{Evolutionary connection between transition objects}  

The connection of these transition objects within their evolutionary progression was 
one of the initial primary objectives of this study, as well as previous studies by \citet{McGregor88a,McGregor88b,
McGregor89} and \citet{Morris96}. 
LHA 115-S 18 (S 18), a supergiant in the SMC, has been the prime example of the possible 
link between the seemingly stable B[e]SGs and the variable LBV stars.  Variability of the star has been found
in a number of different observations.  
The supergiant S 124 similarly shows many characteristics that may indicate it is in an LBV state, however, the stable 
presence of CO emission leaves some ambiguity.  In an exceptional case, \citet{Morris97} found variable CO emission 
in the $K$-band spectrum of the known LBV HR Car, giving further proof of the hazy distinction between B[e]SGs and LBVs. 
The presence of CO emission in LBVs may be observable only at very specific short instances in time, based on the speed 
and physical conditions of the ejected material.  In this vein, we suggest that the cause of the detached
material may be a outburst, or ejection of material from the star, possibly in an S-Dor type event, or due to some 
other physical mechanism such as pulsations.  

As many of the classifications of LBV candidates and B[e]SGs are based on few observations, 
there can be confusion when there are spectral similarities.  For example, the B[e]SG S 65, 
showed no variability in optical spectra, however, recent work by \citet{Oksala12} find that 
CO molecular emission appeared suddenly in the $K$-band spectrum of this star within
a period of 9 months.  This rapid development of spectral features indicates changing conditions in the 
circumstellar environment.  The short time frame of its appearance combined with the large observed velocity of the 
material suggests an outburst of material, reminscent of an S-Dor-type event. 
Further, \citet{Kraus10} determined that this object is located close to the LBV minimum instability strip \citep{Groh}.
But, this is not the only example of possible LBV behavior seen in B[e]SGs.  The spiral arm structure seen in optical spectra
of LHA 120-S 22 by \citet{Aret}, as well as the series of ring material detected around S 127, point to multiple mass ejections 
suggestive of a series of S-Dor-type ejections.  Variability exists in several B[e]SGs, however the study of this group of objects 
suffers from lack of data, with few follow up observations.

\citet{Morris96} inferred that the distinction between B[e]SGs and other transition objects is 
a higher rotational velocity, resulting in a more intensely non-spherical wind, and thus disk, 
a theory supported by the work of \citet{Schulte}. However, a recent study by \citet{Groh} determined that LBVs such as HR Car and 
AG Car are in fact rotating at a significant percentage of their critical velocities.
It is possible that LBVs and B[e]SGs are the same type of stars, with some B[e]SGs not having yet experienced any LBV-type outbursts.

One glaring difference between the group of LBVs and B[e]SGs is the extreme IR excess seen in the color-color diagram 
(see Fig.~\ref{JHHKplot}).  This IR excess indicates that B[e]SGs are surrounded by large amounts of hot dust, 
whereas the LBVs only host marginal amounts of hot dust, but huge amounts of cool dust at far distances.  
This figure shows 2MASS measurements, and is not necessary representative of 
the values at the time of observations.  Since we do not have access to the J and H-band magnitudes at that
epoch, we cannot postulate on any differences, but it is likely that these objects will not have a substantial change in position.
Besides the decreased amount of hot dust and the lack of CO band head emission, the $K$-band spectra of quiescent 
LBVs are otherwise indistinguishable from the B[e]SGs (see Fig.~\ref{CONTspec1}).

\section{Conclusions}

In this paper, a survey of $K$-band spectra from the SINFONI spectrograph has been presented.  Our 25 survey targets 
include LBVs, B[e]SGs, YHGs, and a Peculiar Oe star from the Galaxy, the LMC, and the SMC.  In the following, 
we summarize the main results of this study, as well as make suggestions for future studies.

\begin{itemize}

\item The $K$-band spectra of LBVs in their quiescent or ``hot'' phase are similar in appearance to B[e]SGs.  LBVs in their 
active or ``cool'' phase have $K$-band spectra identical to those observed for YHGs. These similarities include both continuum shape 
and spectral composition, with the exception of CO emission, not found in the spectrum of any of our LBV samples.

\item CO band head emission was detected in 13 of our targets, 10 of which are B[e]SGs, 2 YHGS, and 1 Peculiar Oe star.
For 3 of these stars (MWC 137, LHA 120-S 35, and LHA 115-S 65), this survey presents the initial discovery of 
this feature (the appearance of CO in S 65 has been presented in \citet{Oksala12}).  There appears to be a 
lower luminosity limit, $\log$ L/L$_{\odot}$ = 5.0, below which no CO emission is detected.

\item Model fits to the spectrum reveal the temperature and density of the CO emitting region.
For each object, we find that the CO emitting region must be located in a detached disk or ring 
structure, rather than in a continuous, outflowing disk extending from the stellar surface out to far distances.

\item Each object with visible $^{12}$CO band head emission also shows $^{13}$CO emission, indicating 
evolution past the end of the main sequence.  Based on the enrichment of $^{13}$C, a majority of the stars 
appear to be in a pre-RSG phase, evolving redward across the HRD.  YHGs are post-RSG objects 
preparing to evolve blueward, assuming that stars are able to pass through the 
Yellow Void. B[e]SGs are therefore not the descendants of YHGS.

\item Several of the targets regarded as B[e]SGs appear to be indistinguishable from LBVs, and 
may be regarded as LBV candidates in quiescence.  Variability in several B[e]SGs also suggest that they may be 
candidates as well, further blurring the distinction between these two ambiguous groups.  

\end{itemize}

This survey highlights certain areas where future work can concentrate for these enigmatic groups of stars.  
High resolution spectra of CO
band heads can reveal the kinematics of the disk structure where these molecules are located.  
Simultaneous disk diagnostics, such as [\ion{Ca}{ii}] and [\ion{O}{i}] lines 
and the CO band head emission, are necessary to globally examine the kinematic, temperature, and 
density structure of the circumstellar environments of evolved massive stars.  
To determine the stability of the circumstellar material of these stars over time and determine the 
nature of each object (LBV, B[e]SG, YHG, etc.), further monitoring of the $K$-band spectra (and other 
spectral ranges) of all of these objects is necessary.

\begin{acknowledgements}

The authors would like to express their gratitute to Dr. Adriane Liermann for her assistance in computing
the $K_{s}$-band magnitudes.  This research made use of the NASA Astrophysics Data System (ADS).
M.E.O. and M.K. acknowledge financial support from GA\,\v{C}R under 
grant number 209/11/1198. The Astronomical Institute Ond\v{r}ejov 
is supported by the project RVO:67985815.  M.B.F. acknowledges Conselho 
Nacional de Desenvolvimento Cient\'ifico e Tecnol\'ogico (CNPq-Brazil) 
and Minist\'erio de Ci\^encia, Tecnologia e Inova\c{c}\~ao (MCTI) for a PCI-D grant.
L.C. acknowledges financial support from the 
Agencia de Promoci\'on Cient\'{\i}fica y Tecnol\'ogica (Pr\'estamo BID PICT 2011/0885), 
PIP 0300 CONICET, and the Programa de Incentivos G11/109 of the Universidad 
Nacional de La Plata, Argentina.  Financial support from International Cooperation 
of the Czech Republic (M\v{S}MT, 7AMB12AR021) and Argentina (Mincyt-Meys, ARC/11/10) 
is acknowledged.  M.F.M. is a research fellow of the Universidad 
Nacional de La Plata, Argentina.  

\end{acknowledgements}

\end{document}